\documentclass{article}

\usepackage{arxiv}
\usepackage[square,comma,numbers]{natbib}
\usepackage{graphicx}
\usepackage{subcaption}
\usepackage{caption}
\usepackage{amsmath}
\usepackage{amsfonts}
\usepackage{amssymb}
\usepackage{bbm}
\usepackage{listings}
\usepackage{wrapfig}
\usepackage{mathtools}
\usepackage{algorithm}
\usepackage{algpseudocode}

\newtheorem{Thm}{Theorem}
\usepackage{bm}

\usepackage[utf8]{inputenc} 
\usepackage[T1]{fontenc}    
\usepackage{hyperref}       
\usepackage{url}            
\usepackage{booktabs}       
\usepackage{amsfonts}       
\usepackage{nicefrac}       
\usepackage{microtype}      
\usepackage{lipsum}

\title{Innovative And Additive Outlier Robust Kalman Filtering With A Robust Particle Filter}

\author{
  Alexander T.\ M.\ Fisch\\
  Lancaster University\\
  Lancaster, United Kingdom\\
   \And
  Idris A.\ Eckley \\
  Lancaster University\\
  Lancaster, United Kingdom\\
  \AND
  Paul Fearnhead \\
  Lancaster University\\
  Lancaster, United Kingdom\\
}

\begin{document}
\maketitle

\begin{abstract}
    In this paper, we propose CE-BASS, a particle mixture Kalman filter which is robust to both innovative and additive outliers, and able to fully capture multi-modality in the distribution of the hidden state. Furthermore, the particle sampling approach re-samples past states, which enables CE-BASS to handle innovative outliers which are not immediately visible in the observations, such as trend changes. The filter is computationally efficient as we derive new, accurate approximations to the optimal proposal distributions for the particles. The proposed algorithm is shown to compare well with existing approaches and is applied to both machine temperature and server data. 
\end{abstract}


\section{Introduction And Literature Review}

Anomaly detection is an area of considerable importance and has been subject to increasing attention in recent years. Comprehensive reviews of the area can be found in \cite{chandola2009anomaly, pimentel2014review}. The field's growing importance arises from the increasing range of applications to which anomaly detection lends itself: from fraud prevention \citep{chandola2009anomaly,pimentel2014review}, to fault detection  \citep{chandola2009anomaly,pimentel2014review}, and even the detection of exoplanets \citep{fisch2018linear}. More recently, the emergence of internet of things and the ubiquity of sensors has led to emergence of the online detection of anomalies as an important statistical challenge.

Kalman filters \cite{Kalman1960} provide a convenient framework to detect anomalies within a streaming data context. In particular, they can be updated in a fully online fashion at a fixed computational cost. At each time point, Kalman filters also provide an estimate both for the expectation and variance of the next observation. These can be used to determine whether that observation is anomalous or not. However, the major drawback of Kalman filters is their lack of robustness to outliers: once the filter has encountered an outlier, it will often produce inaccurate predictions for many future time points.

The anomaly detection literature distinguishes between two types of outliers. The first are additive outliers, sometimes referred to as observational outliers \citep{gandhi2009robust}, which affect the observational noise only. The other type of outliers are the innovative, or process \citep{huang2017novel}, outliers. These affect the updates of the hidden states. 
In practice, both have a similar effect on the next observation, but quite different effects on subsequent observations. Moreover, some innovative outliers cannot be detected immediately as their influence on the observations is only noticeable after, or over, a period of time. 


A range of robust Kalman filters has been proposed to date. Many side-step the problem of distinguishing between the two outlier types. By far the largest class of filters aims to be robust against heavy tailed additive outliers. Examples of such filters include \cite{ting2007learning, agamennoni2011outlier}, which assume $t$-distributed additive noise and perform inference using variational Bayes, \cite{ruckdeschel2014robust}, who use Huberised residuals, and \cite{chang2014robust} inflate the noise covariance matrix whenever an outlier is encountered. A few filters have also been developed with the aim of achieving robustness against innovative outliers \citep{ruckdeschel2014robust}. The problem with such filters is that they exacerbate the shortcomings of the Kalman filter when they encounter the other type of anomaly: additive outlier robust Kalman filters, for example, update their hidden states even less than the classical Kalman filter when encountering innovative outliers. 

In principle, it seems straightforward to combine the ideas of these two types of robust Kalman filter. One body of literature proposes to use Huberisation of both innovative and additive residuals \citep{gandhi2009robust,chang2014robust}. Others \citep{huang2017novel,huang2019novel} have modelled both additive and innovative outliers using $t$-distributions, by imposing Wishart priors on the precision matrix of both the innovations and additions and maintaining the posterior by using variational Bayes approaches. The issue with these filters comes from how they approximate the filtering distribution of the state. Both return uni-modal posteriors after encountering an anomaly. This is a shortcoming given that the posterior after an anomaly is likely to be multi-modal: if the outlying observation was caused by an additive anomaly, the state will be close to the prior, whereas if it was caused by an innovative anomaly, the state would be far from it. 

The ideal approach to constructing a robust filter would be to model the possibility of outliers in both the observation and system noise, and then use a filter algorithm that attempts to calculate, or approximate, the true filtering distribution for the model. An early attempt to do this was the spline based approach \cite{kitagawa1987non}, but the computational complexity increases very quickly with the number of dimensions and such a filter becomes impracticable when the state dimension is greater than 3. 
As a result we consider using particle filters \citep{gordon1993novel,fearnhead2018particle}. These are able to produce Monte Carlo approximations to the filtering distribution for an appropriate model that allows for outliers, and, in principle, can work even if the filtering distribution is multi-modal. However the Monte Carlo error of standard implementations of the particle can be prohibitively large \citep{chang2014robust}. 

In this paper, we develop an efficient particle filter by using a combination of Rao-Blackwellisation and well-designed proposal distributions. The idea of Rao-Blackwellisation is to integrate out part of the state so that the particle filter approximates the filtering distribution of a lower-dimensional projection of the state. In our application this projection is whether each component of the additive and innovative noise is an outlier, and if it is how much the variance of the noise has been inflated. Conditional on this information, the state space model becomes linear-Gaussian and we can implement a Kalman Filter to calculate exactly the conditional filtering distribution, while being able to fully capture multi modal posteriors. This idea is similar to that which underpins the Mixture Kalman Filter \citep{chen2000mixture}. 

Whilst Rao-Blackwellisation improves the Monte Carlo accuracy of the filter, such a filter can still have the shortcomings noted by \cite{chang2014robust} and perform poorly without good proposal distributions for the information we condition on. One of the main contributions of this work is a proposal distribution that accurately approximates the conditional distribution of the variance inflation for each component of the noise, and hence approximates the optimal proposal distribution \citep{pitt1999filtering}. As a result of this proposal, we find that accurate results can be obtained even with only a few particles. 

Another important challenge addressed by this paper is that certain innovative outliers can not immediately be detected. An innovative outlier in a latent trend component for instance can cause a trend changes which may only become apparent -- i.e.\ produce a visible outlier in the observations -- many observations after the innovative outlier in the trend occurred. It is nevertheless important to capture such outliers as they can affect a potentially unlimited number of observations to come. The proposed particle filter includes the possibility to back-sample the variance inflation particles in light of more recent observations, which enables it to capture these important anomalies.

The remainder of this paper is organised as follows: We discuss our robust noise model, consisting of a mixture distribution of Gaussian noise, representing typical behaviour, and heavy tailed noise, representing atypical behaviour, for both the additive (observational) and innovative (system) noise process in Section \ref{sec:Model}. The model is shown to be very similar to that considered by \cite{huang2019novel}. We then introduce the proposal distribution for the scale of the noise in Section \ref{sec:Particle_Filter}, before extending it to anomalies which are not immediately identifiable in Section \ref{sec:Backsampling}. The proposed filter is compared to others in Section \ref{sec:Simulation} and applied to router data and a benchmark machine temperature data-set in Section \ref{sec:Application}. The proposed methodology, which we call Computationally Efficient Bayesian Anomaly detection by Sequential Sampling (CE-BASS) has been implemented in the the \texttt{R} package \texttt{RobKF} available from \texttt{https://github.com/Fisch-Alex/Robkf}. Derivations of theoretical results and complete pseudocode are available in the appendix.

\section{Model And Examples}\label{sec:Model}
Throughout this paper, we will consider inference about a latent state, $\textbf{X}_t$, through partial observations, $\textbf{Y}_t$, modelled as
\begin{align}\label{eq:Main}
\begin{split}
\textbf{Y}_t &= \textbf{C} \textbf{X}_t +   \textbf{V}_t^{\frac{1}{2}}\bm{\Sigma}_A^{\frac{1}{2}}\bm{\epsilon}_t,  \\
\textbf{X}_t &= \textbf{A} \textbf{X}_{t-1} +   \textbf{W}_t^{\frac{1}{2}}\bm{\Sigma}_I^{\frac{1}{2}}\bm{\nu}_t.
\end{split}
\end{align}
Here the additive noise, $\bm{\epsilon}_t \in \mathbb{R}^p$, and the innovations $\bm{\nu}_t \in \mathbb{R}^q$ are both i.i.d.\ standard multivariate Gaussian. The diagonal matrices $\bm{\Sigma}_A$ and $\bm{\Sigma}_I$ denote the covariance of the additive and innovation noise respectively. The diagonal matrices $\textbf{V}_t$ and $\textbf{W}_t$ are used to capture additive and innovative outliers respectively, with large diagonal entries of $\textbf{V}_t$ corresponding to additive outliers and large diagonal entries of $\textbf{W}_t$ corresponding to innovative outliers. The classical Kalman model is recovered by setting $\textbf{W}_t=\textbf{I}$ and $\textbf{V}_t=\textbf{I}$ for all times $t$. 

\begin{figure}
	\begin{subfigure}[b]{0.49\linewidth}
		\centering
		\includegraphics[width=0.9\linewidth]{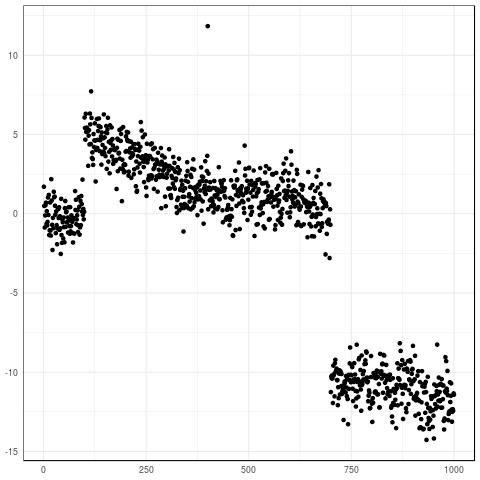}
		\caption{Random walk}\label{fig:rw_ex} 
	\end{subfigure}
	\begin{subfigure}[b]{0.49\linewidth}
		\centering
		\includegraphics[width=0.9\linewidth]{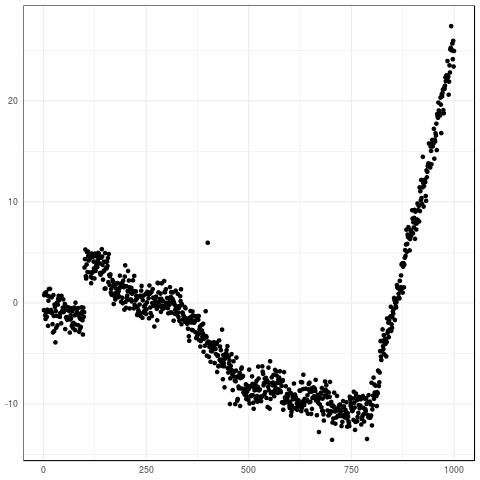}
		\caption{Random walk with trends}	\label{fig:rwandt_ex} 
	\end{subfigure}
	\caption{Two examples of time series which are realisations of outlier infested Kalman models. (a) was simulated using the setup defined in Equation \eqref{eq:RW}, with $\sigma_A = 1$, $\sigma_I = 0.1$, and outliers defined by $W_{100} = 3600$, $V_{400} = 100$, and $W_{700} = 10000$. Conversely (b) second example was simulated using the model defined in Equation \eqref{eq:RWand_T} using $\sigma_A = 1$, $\sigma_I^{(1)} = 0.1$, $\sigma_I^{(2)} = 0.01$ and outliers defined by $W^{(1)}_{100} = 3600$, $V_{400} = 100$, and $W^{(2)}_{700} = 40000$.}
\end{figure}

The model in Equation \eqref{eq:Main} can be used to model a range of time series behaviours. We will use the following two examples throughout the paper:

 \textbf{Example 1}: The random walk model with both changepoints and outliers, similar to the problem considered by \cite{fearnhead2019changepoint}. It can be formulated as 
\begin{align}\label{eq:RW}
Y_t = X_t + V_t^{\frac{1}{2}}\sigma_A\epsilon_t, \;\;\;\;\;\;\;\;\;\;\;\;\;
X_t = X_{t-1} + W_t^{\frac{1}{2}}\sigma_I\nu_t.
\end{align}
Here atypically large values of $V_t$ correspond to outliers, whilst atypically large values of $W_t$ correspond to changes. A realisation of this model can be found in Figure \ref{fig:rw_ex}. 

\textbf{Example 2}: A time series with changes in trend, level shifts, as well as outliers, similar to the model considered by \cite{maeng2019detecting}. It can be formulated as 
\small
\begin{equation}\label{eq:RWand_T}
\begin{split}
Y_t = X_t^{(1)} + V_t^{\frac{1}{2}}\sigma_A\epsilon_t  \;\;\;\;\;\;\;\;\;\;\;\;\;
X_t^{(1)} &= X_{t-1}^{(1)} + X_{t-1}^{(2)} + \left(W_t^{(1)}\right) ^{\frac{1}{2}}\sigma_I^{(1)}\nu_t^{(1)},\\
  X_t^{(2)} &= X_{t-1}^{(2)} + \left(W_t^{(2)}\right) ^{\frac{1}{2}}\sigma_I^{(2)}\nu_t^{(2)},
\end{split}
\end{equation}
\normalsize
with the first component of the hidden state denoting the current position and the second indicating the trend. Here, outliers are modelled by large values of $V_t$ whilst level shift and changes in trend are modelled by atypically large values of $W_t^{(1)}$ and $W_t^{(2)}$ respectively. A realisation of this model can be found in Figure \ref{fig:rwandt_ex}. 

A key feature of this second model is that an outlier in the trend component, $X_t^{(2)}$, may only become detectable many observations after the outlier -- this challenging issue mentioned in the introduction is addressed via the methods in Section \ref{sec:Backsampling}. A wide rage of other commonly used time series features, such as auto-correlation, moving averages, etc.\ can be incorporated in the model. 

To infer the locations of anomalies we use the model
\begin{equation}\label{eq:Noise}
    \textbf{V}_t^{(i,i)} = 1 + \lambda_t^{(i)} \frac{1}{\tilde{\textbf{V}}_t^{(i,i)}} \;\;\;\;\;\;\;\;\;\textbf{W}_t^{(j,j)} = 1 + \gamma_t^{(j)} \frac{1}{\tilde{\textbf{W}}_t^{(j,j)}}
\end{equation}
for $1 \leq i \leq p$ and $1 \leq j \leq q$. The random variables $\lambda_t^{(i)} \sim Ber(r_i)$ and $\gamma_t^{(j)} \sim Ber(s_j)$ are indicators that determine whether an anomaly is present or not for $1 \leq i \leq p$ and $1 \leq j \leq q$ respectively. For additional interpretability, we impose that at most one anomaly is present at any given time $t$, and define $r_i$ and $s_j$ to be the probabilities that $\lambda_t^{(i)}=1$ and $\gamma_t^{(j)}=1$ respectively. The inverse scale, or precision, of an anomaly (if present) is given by the random variables $\tilde{\textbf{V}}_t^{(i,i)} \sim \tilde{\sigma_i}\Gamma(a_i,a_i)$ and $\tilde{\textbf{W}}_t^{(j,j)} \sim \hat{\sigma_j}\Gamma(b_j,b_j)$ for $1 \leq i \leq p$ and $1 \leq j \leq q$ respectively.

The proposed model bears similarities to the model used by \cite{huang2019novel}. Both use a mixture of Gaussian and heavy tailed noise. The main difference is that the anomalous behaviour is characterised by noise which is the sum of a Gaussian and a $t$-distribution in our model as opposed to just a $t$-distribution in the model used by \cite{huang2019novel}. This ensures that anomalies coincide with strictly greater noise and makes the result more interpretable. In practice, however, the noise distribution considered in this paper and in \cite{huang2019novel} are likely to be of very similar shape.

\section{Particle Filter}\label{sec:Particle_Filter}

We now turn to filtering the model defined by Equations \eqref{eq:Main} and \eqref{eq:Noise}. The main feature we exploit is the fact that if we knew the value of $(\textbf{V}_t,\textbf{W}_t)$ at all times $t$, we could just run the classical Kalman filter over the data. Consequently, our approach will consist of sampling particles for $(\textbf{V}_t,\textbf{W}_t)$, conditional on which the classical Kalman update equations for the hidden state $\textbf{x}_t$ can be used. This approach, very similar to the mixture Kalman filter \citep{chen2000mixture,fearnhead2003line} is summarised by the pseudocode in Algorithm \ref{alg:Basic}. 

For each time, $t$, the code loops over the existing particles, $(\textbf{V}_{t},\textbf{W}_{t})$, and simulates $M'$ descendants for each of them in step 4. They are stored in a set of candidate particles. If we have $N$ particles at time $t$, keeping all candidates would produce $NM'$ particles at time $t+1$. To avoid growing the number of particles exponentially with $t$, Step 7 resamples the candidates to keep just $N$ particles. The filtering distribution for each of these particles is then calculated using the Kalman Filter updates in step 10.


\begin{algorithm}
	\caption{Basic Particle Filter (No Back-sampling)}
	\label{alg:Basic}
	\begin{footnotesize}
		\begin{tabular}[h]{ll}
			{\bf Input:} & An initial state estimate $(\bm{\mu}_0,\bm{\Sigma}_0)$ \\ & A number of descendants, $M'=M(p+q)+1$ \\ & A number of particles to be maintained, $N$.
			 \\ & A stream of observations $\textbf{Y}_1,\textbf{Y}_2,...$ \\ {\bf Initialise:} & Set $Particles(0) = \{(\bm{\mu}_0,\bm{\Sigma}_0)\}$ 
		\end{tabular}

		\begin{algorithmic}[1]
			\For{$t \in \mathbb{N}^+ $}
			\State $Candidates \gets \{\}$ 
			\For{$(\bm{\mu},\bm{\Sigma}) \in Particles(t-1)$}
			\State $(\textbf{V},\textbf{W},prob) \gets \text{Sample\_Particles}(M',\bm{\mu},\bm{\Sigma},\textbf{Y}_t,\textbf{A},\textbf{C},\bm{\Sigma}_A,\bm{\Sigma}_I)$
			\State $Candidates \gets Candidates \cup \{(\bm{\mu},\bm{\Sigma},\textbf{V},\textbf{W},prob)\}$ 
			\EndFor
			\State $Descendants \gets \text{Subsample}(N,Candidates)$
			\State $Particles(t) \gets \{\}$
			\For{$(\bm{\mu},\bm{\Sigma},\textbf{V},\textbf{W},prob) \in Descendants $}
			\State $(\bm{\mu}_{new},\bm{\Sigma}_{new}) \gets  \text{KF\_Upd}(\textbf{Y}_t,\bm{\mu},\bm{\Sigma},\textbf{C},\textbf{A},\textbf{V}^{1/2}\bm{\Sigma}_A,\textbf{W}^{1/2}\bm{\Sigma}_I)$
			\State $Particles(t) \gets Particles(t) \cup \{(\bm{\mu}_{new},\bm{\Sigma}_{new}) \}$
			\EndFor
			\EndFor
		\end{algorithmic}
	\end{footnotesize}
\end{algorithm}

The main challenge in the above approach consists of selecting a good sampling procedure for the particles. Whilst it may be a natural choice to sample particles $(\textbf{V}_{t+1},\textbf{W}_{t+1})$ from their prior distribution, this is not suitable for the problem considered in this paper. In particular, this sampling procedure would not be robust to outliers: the stronger an anomaly was, the less likely we would be to sample a particle with an appropriate value of $(\textbf{V}_{t+1},\textbf{W}_{t+1})$, as discussed by \cite{chang2014robust}. 

\normalsize
Adopting ideas from \cite{pitt1999filtering} and \cite{arulampalam2002tutorial}, we overcome the above challenge by sampling particles from an approximation to the conditional distribution of
$(\textbf{V}_{t+1},\textbf{W}_{t+1})$ given observation $\textbf{Y}_{t+1}$. Denote the model's prior distribution for $(\textbf{V}_{t+1},\textbf{W}_{t+1})$ in \eqref{eq:Noise} by $\pi_0(\cdot)$. The conditional distribution $\pi(\textbf{W}_{t+1},\textbf{V}_{t+1}|\textbf{Y}_{t+1})$ for the descendants of a particle whose filtering distribution for $\textbf{x}_{t}$ is $N(\bm{\mu},\bm{\Sigma})$ is then proportional to 
\begin{equation*}
\pi_0(\textbf{W},\textbf{V}) \mathcal{L}\left(\textbf{Y},\textbf{C}\textbf{A}, \textbf{C}\textbf{A}\bm{\Sigma}\textbf{A}^T\textbf{C}^T + \bm{\Sigma}_A \textbf{V}+ \textbf{C} \bm{\Sigma}_I\textbf{W}\textbf{C}^T\right).
\end{equation*} 
Here we have dropped time indices for convenience, and $\mathcal{L}\left(\textbf{x}, \bm{\mu}, \bm{\Sigma}\right)$ denotes the likelihood of an observation $\textbf{x}$ under a $N(\bm{\mu}, \bm{\Sigma})$-model. Since at most one component is anomalous, we can re-write this as a sum over which, if any, component is anomalous
\small
\begin{align*}
\mathbb{I}_{\left\{ \textbf{W}=\textbf{I},\textbf{V}=\textbf{I} \right\}}\pi(\textbf{I},\textbf{I}|\textbf{Y}) + \sum_{j=1}^q \mathbb{I}_{\left\{ \textbf{W}=\textbf{I} + \frac{\textbf{I}^{(j)}}{\tilde{\textbf{W}}^{(j,j)}},\textbf{V}=\textbf{I} \right\}}\hat{\pi}_j\left(\tilde{\textbf{W}}^{(j,j)}\right) + 
\sum_{i=1}^p \mathbb{I}_{\left\{ \textbf{W}=\textbf{I},\textbf{V}=\textbf{I} + \frac{\textbf{I}^{(i)}}{\tilde{\textbf{V}}^{(i,i)}} \right\}}\tilde{\pi}_i\left(\tilde{\textbf{V}}^{(i,i)}\right).
\end{align*}
\normalsize
Here, we use the shorthand
\footnotesize
\begin{equation*}
 \tilde{\pi}_i\left(\tilde{\textbf{V}}^{(i,i)}\right) = \pi\left( \textbf{I},\textbf{I} + \frac{\textbf{I}^{(i)}}{\tilde{\textbf{V}}^{(i,i)}}|\textbf{Y} \right) \end{equation*} \normalsize and \footnotesize \begin{equation*}
 \hat{\pi}_j\left(\tilde{\textbf{W}}^{(j,j)}\right) = \pi\left( \textbf{I} + \frac{\textbf{I}^{(j)}}{\tilde{\textbf{W}}^{(j,j)}} ,\textbf{I}|\textbf{Y} \right).
\end{equation*}
\normalsize

Since the target distribution $\pi(\textbf{W},\textbf{V}|\textbf{Y})$ is intractable, we construct an approximation to it, which we denote $q(\textbf{W},\textbf{V}|\textbf{Y})$, and use this as our proposal distribution. This proposal  is proportional to
\small
\begin{align*}
\mathbb{I}_{\left\{ \textbf{W}=\textbf{I},\textbf{V}=\textbf{I} \right\}}\beta_0 + \sum_{j=1}^q \mathbb{I}_{\left\{ \textbf{W}=\textbf{I} + \frac{\textbf{I}^{(j)}}{\tilde{\textbf{W}}^{(j,j)}},\textbf{V}=\textbf{I} \right\}} \hat{\beta}_j \hat{q}_j\left(\tilde{\textbf{W}}^{(j,j)}\right) + 
\sum_{i=1}^p \mathbb{I}_{\left\{ \textbf{W}=\textbf{I},\textbf{V}=\textbf{I} + \frac{\textbf{I}^{(i)}}{\tilde{\textbf{V}}^{(i,i)}} \right\}}\tilde{\beta}_i\tilde{q}_i\left(\tilde{\textbf{V}}^{(i,i)}\right).
\end{align*}
\normalsize
Clearly, there is no benefit in simulating multiple identical descendants, so we wish to sample precisely one dependent that corresponds to no outliers. To do this, and also to have the same number of descendant particles for each possible type of outlier, we set $\beta_0 = \frac{1}{1+M(p+q)}$,  $\tilde{\beta}_i = \frac{M}{1+M(p+q)}$,  and $\hat{\beta}_j = \frac{M}{1+M(p+q)}$, and use stratified subsampling as in \cite{fearnhead2003line}. This leads to $M'=M(p+q)+1$ total descendants per particle, $M$ for each of the $p$ additive and $q$ innovative outliers, and one for no outlier. Each of these particles is then given a weight proportional to 
\begin{equation*}
    \frac{\pi(\textbf{W}_{t+1},\textbf{V}_{t+1}|\textbf{Y}_{t+1})}{q(\textbf{W}_{t+1},\textbf{V}_{t+1}|\textbf{Y}_{t+1})}.
\end{equation*}

The main challenge now consists of obtaining proposal distributions $\tilde{q}_i(\cdot)$ for $1 \leq i \leq p$ and  $\hat{q}_j(\cdot)$ for $1 \leq j \leq q$ that provide good approximations to the conditional posteriors which are proportional to $\tilde{\pi}_i(\cdot)$ and $\hat{\pi}_j(\cdot)$ respectively. 
In the next subsection, we therefore derive proposal distributions that provide leading order approximations to the conditional posteriors. To simplify notation, we define the predictive variance 
$\hat{\bm{\Sigma}} = \textbf{C}\textbf{A}\bm{\Sigma}\textbf{A}^T\textbf{C}^T + \bm{\Sigma}_A + \textbf{C} \bm{\Sigma}_I\textbf{C}^T$
and use it throughout the remainder of this paper. We also begin by assuming that $\textbf{C}$ contains no $0$-columns. The proposal introduced in the following subsection also forms the basis of back-sampling introduced in Section \ref{sec:Backsampling}, which allows to relax this on $\textbf{C}$.

\subsection{Proposal Distributions}\label{sec:Props}

For $1 \leq i \leq p$, we would like the proposal distribution $\tilde{q}_i\left(\tilde{\textbf{V}}^{(i,i)}\right)$ for the precision, $\tilde{\textbf{V}}^{(i,i)}$, to be as close as possible to  $\tilde{\pi}_i\left(\tilde{\textbf{V}}^{(i,i)}\right)$ or, equivalently, proportional to
\footnotesize
\begin{equation*} f_i\left(\tilde{\textbf{V}}^{(i,i)}\right)
\frac{
	\exp \left( 
	-\frac{1}{2} \left(\textbf{Y}-\textbf{C}\textbf{A}\bm{\mu}\right)^T
	\left( \hat{\bm{\Sigma}} +  \frac{\bm{\Sigma}_A^{(i,i)}}{\tilde{\textbf{V}}^{(i,i)}} \textbf{I}^{(i)}\right)^{-1} 
	\left(\textbf{Y}-\textbf{C}\textbf{A}\bm{\mu}\right)
	\right)
}{\sqrt{\left|\hat{\bm{\Sigma}} +  \frac{\bm{\Sigma}_A^{(i,i)}}{\tilde{\textbf{V}}^{(i,i)}} \textbf{I}^{(i)} \right|}},
\end{equation*}
\normalsize
where $f_i()$ denotes the PDF of the $\tilde{\sigma}_i\Gamma(a_i,a_i)$-distributed prior of $\tilde{\textbf{V}}^{(i,i)}$. 

It should be noted that the intractable terms, 
\begin{equation}\label{eq:annoyingterms}
   \left|\hat{\bm{\Sigma}} +  \frac{\bm{\Sigma}_A^{(i,i)}}{\tilde{\textbf{V}}^{(i,i)}} \textbf{I}^{(i)} \right| \;\;\;\;\;\;\; \text{and} \;\;\;\;\;\;\; \left( \hat{\bm{\Sigma}} +  \frac{\bm{\Sigma}_A^{(i,i)}}{\tilde{\textbf{V}}^{(i,i)}} \textbf{I}^{(i)}\right)^{-1}
\end{equation}
can both be expanded using the matrix determinant lemma and the Sherman Morrison formula respectively, as they are rank 1 updates of a determinant and inverse respectively. 
Indeed, by the matrix determinant lemma, 
\small
\begin{equation*}
   \left|\hat{\bm{\Sigma}} +  \frac{\bm{\Sigma}_A^{(i,i)}}{\tilde{\textbf{V}}^{(i,i)}} \textbf{I}^{(i)} \right| =  \frac{\left|\hat{\bm{\Sigma}}\right| }{\tilde{\textbf{V}}^{(i,i)}}\left( 1 +  \bm{\Sigma}_A^{(i,i)}\left(\hat{\bm{\Sigma}}^{-1} \right)^{(i,i)} + O\left(\tilde{\textbf{V}}^{(i,i)} \right) \right),
\end{equation*}
\normalsize
the leading order term is conjugate to the prior of $\tilde{\textbf{V}}^{(i,i)}$. Moreover, by the  Sherman Morrison formula the second term in Equation \eqref{eq:annoyingterms} is equal to
\small
\begin{equation*}
\hat{\bm{\Sigma}}^{-1} - \hat{\bm{\Sigma}}^{-1} \textbf{I}^{(i)}  \hat{\bm{\Sigma}}^{-1} \left[\frac{1}{\left(\hat{\bm{\Sigma}}^{-1}\right)^{(i,i)}} -  \left(\frac{1}{\left(\hat{\bm{\Sigma}}^{-1}\right)^{(i,i)}}\right)^2\frac{\tilde{\textbf{V}}^{(i,i)}}{\bm{\Sigma}_A^{(i,i)}}\right], 
\end{equation*}
\normalsize
up to $ O \left( \left(\tilde{\textbf{V}}^{(i,i)} \right)^2 \right) $. Crucially, the first two terms are constant in $\tilde{\textbf{V}}^{(i,i)}$, while the third is linear in $\tilde{\textbf{V}}^{(i,i)}$ and therefore returns a term which is conjugate to the prior of $\tilde{\textbf{V}}^{(i,i)} $. Furthermore, we are most concerned about accurately sampling the particle when an anomaly occurs in the $i$th component, which happens when the precision, $\tilde{\textbf{V}}^{(i,i)}$, and the higher order terms, become small. 

Keeping only the leading order terms in the determinant and the exponential term results in the proposal distribution 
\small
	\begin{equation*}
	\tilde{\textbf{V}}^{(i,i)}\sim \tilde{\sigma}_i\Gamma\left(a_i + \frac{1}{2},a_i + \frac{\tilde{\sigma}_i}{2\bm{\Sigma}_A^{(i,i)}}\left( \frac{\left(\hat{\bm{\Sigma}}^{-1}\right)^{(i,:)} \left(\textbf{Y}-\textbf{C}\textbf{A}\bm{\mu}\right) }{\left(\hat{\bm{\Sigma}}^{-1}\right)^{(i,i)}}\right)^2\right)
	\end{equation*}
\normalsize
for $\tilde{\textbf{V}}^{(i,i)}$.	More detailed derivations, including the associated weight are given by Theorem 1 in the appendix. This proposal has the property that as the observed anomaly in the $i$th component becomes larger, i.e.\ as 
\begin{equation*}
\frac{1}{\bm{\Sigma}_A^{(i,i)}}\left( \frac{\left(\hat{\bm{\Sigma}}^{-1}\right)^{(i,:)} \left(\textbf{Y}-\textbf{C}\textbf{A}\bm{\mu}\right) }{\left(\hat{\bm{\Sigma}}^{-1}\right)^{(i,i)}}\right)^2
\end{equation*} 
increases, the mean of the proposal for $\tilde{\textbf{V}}^{(i,i)}$ diverges from the prior mean and behaves asymptotically like
\begin{equation*}
(2a_i+1)\bm{\Sigma}_A^{(i,i)} \left( \frac{\left(\hat{\bm{\Sigma}}^{-1}\right)^{(i,i)}}{\left(\hat{\bm{\Sigma}}^{-1}\right)^{(i,:)} \left(\textbf{Y}-\textbf{C}\textbf{A}\bm{\mu}\right) }\right)^2.
\end{equation*}
Consequently, the variance and the squared residual will be on the same scale, thus achieving computational robustness. 

A very similar approach can be used to obtain a proposal distribution $\hat{q}_j \left( \tilde{\textbf{W}}^{(j,j)} \right)$ which provides a leading order approximation for the distribution proportional to $\pi \left( \textbf{I} + \frac{1}{\tilde{\textbf{W}}^{(j,j)}} \textbf{I}^{(j)} ,\textbf{I} |\textbf{Y} \right)$. The proposal consists of sampling 
\footnotesize
\begin{equation*}
	\tilde{\textbf{W}}^{(j,j)} \sim \hat{\sigma_j}\Gamma\left(b_j + \frac{1}{2},b_j + \frac{\hat{\sigma_i}}{2\bm{\Sigma}_I^{(j,j)}}\left( \frac{\left(\textbf{C}^T\right)^{(j,:)}\hat{\bm{\Sigma}}^{-1} \left(\textbf{Y}-\textbf{C}\textbf{A}\bm{\mu}\right) }{\left(\textbf{C}^T\hat{\bm{\Sigma}}^{-1}\textbf{C}\right)^{(j,j)}}\right)^2\right)
\end{equation*} 
\normalsize
and is of very similar form to the proposal distribution for particles with an additive outlier and well defined if $\textbf{C}$ has no $\textbf{0}$-columns. Further details, including the associated weight, are given in Theorem 2 in the appendix. Like the proposal distribution for particles with an additive anomaly this proposal is computationally robust: it ensures that the squared residual and the variance will be on the same scale as the anomaly in the $j$th innovative component becomes stronger. 

Finally, the ``proposal" for particles without anomalies consists of deterministically setting $\textbf{V} = \textbf{I}$ and $\textbf{W} = \textbf{I}$. The weight associated with this particle is proportional to the likelihood, the closed form of which is given in Theorem 3 in the appendix.

\subsection{Choices of Parameters}\label{sec:weights}

The choice of hyper-parameters, particularly $\hat{\sigma_i}$ and $\tilde{\sigma_i}$, has a significant effect of the performance of the proposed filter. One reason for this is that an outlier observation could be the result of either an additive or an innovative outlier. It may be that the root cause can only be determined  after further observations are made. Thus, we wish to choose hyper-parameters in such a way as to ensure that observed anomalies, which are equally well explained by different classes of anomalies, are given similar importance weights. The following result describes such a choice: 
\setcounter{Thm}{3}
\begin{Thm}\label{Thm:Weights}
	Let the prior for the hidden state $\textbf{X}_{t}$ be $N(\bm{\mu},\bm{\Sigma})$ and an observation $\bm{Y}_{t+1} := \bm{Y}$ be available. When 
\begin{equation*}
\tilde{\sigma}_i = \Sigma_A^{(i,i)} \left(\hat{\bm{\Sigma}}^{-1}\right)^{(i,i)}
\;\;
\text{and} \;\;
\hat{\sigma}_j =\Sigma_I^{(j,j)} \left(\textbf{C}^T\hat{\bm{\Sigma}}^{-1}\textbf{C}\right)^{(j,j)},
\end{equation*}
and $a_1 = ... = a_p = b_1 = ... = b_q = c$, the weights of additive and innovative anomalies are asymptotically proportional to 
\begin{equation*}
\frac{c^{c}\frac{1}{M}r_i\frac{\Gamma(c+\frac{1}{2})}{\Gamma(c)} 
	\exp \left( 
	\frac{1}{2} \delta ^ 2 
	\right)
}{\left( 
	\frac{\delta^2}{2}
	\right)^{c}
} 
\;\;
\text{and}
\;\;
\frac{c^{c}\frac{1}{M}s_j\frac{\Gamma(c+\frac{1}{2})}{\Gamma(c)} 
	\exp \left( 
	\frac{1}{2} \delta ^ 2 
	\right)
}{\left( 
	\frac{\delta^2}{2}
	\right)^{c}
} 
\end{equation*}
when 
\small
\begin{equation*}
\textbf{Y}-\textbf{CA}\bm{\mu} = \frac{\delta \textbf{e}_i}{ \sqrt{\left( \hat{\bm{\Sigma}}^{-1}\right)^{(i,i)}} }
\;\;
\text{and} \;\;
\textbf{Y}-\textbf{C}\textbf{A}\bm{\mu} = \frac{\delta\textbf{C}^{(:,j)}}{\sqrt{\left(\textbf{C}^T\hat{\bm{\Sigma}}^{-1}\textbf{C}\right)^{(j,j)}}},
\end{equation*}
\normalsize
respectively, as $\delta \rightarrow \infty$
\end{Thm}

The above choice of hyper-parameters therefore leads to all components being given equal asymptotic importance weight under an anomaly they are able to account for. I.e.\ one which satisfies $\frac{\textbf{C}^{(:,j)}}{\sqrt{\left(\textbf{C}^T\hat{\bm{\Sigma}}^{-1}\textbf{C}\right)^{(j,j)}}}\delta = \textbf{Y}-\textbf{CA}\bm{\mu} = \frac{\delta \textbf{e}_i}{ \sqrt{\left( \hat{\bm{\Sigma}}^{-1}\right)^{(i,i)}} }  $. Setting all the $a_i$s and $b_j$s to the same constant is advisable due to the fact that the convolution of two $t$-distributions whose means drift further and further apart yields two stable, i.e.\ non-vanishing modes if and only if they have the same scale parameter. 

While, $\hat{\bm{\Sigma}}^{-1}$ is not fixed but time dependent, it nevertheless converges to a limit under an observable Kalman filter model. In practice, we therefore use this limit to set $\tilde{\sigma}_i$ and $\hat{\sigma}_j$.

\subsection{Example 1 - revisited}

\begin{figure}
	\begin{subfigure}[b]{0.32\linewidth}
		\centering
		\includegraphics[width=0.9\linewidth]{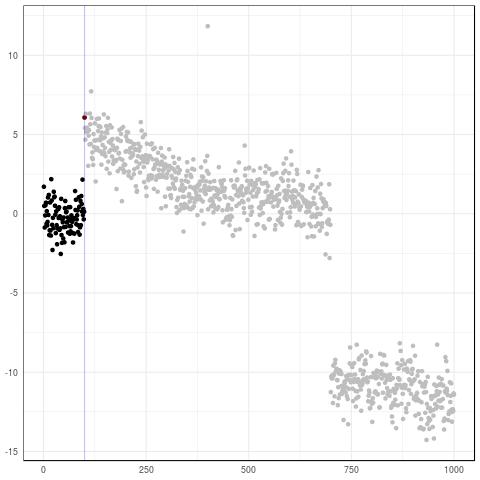}
		\caption{t=100} \label{fig:rw_ex_100} 
	\end{subfigure}
	\begin{subfigure}[b]{0.32\linewidth}
		\centering
		\includegraphics[width=0.9\linewidth]{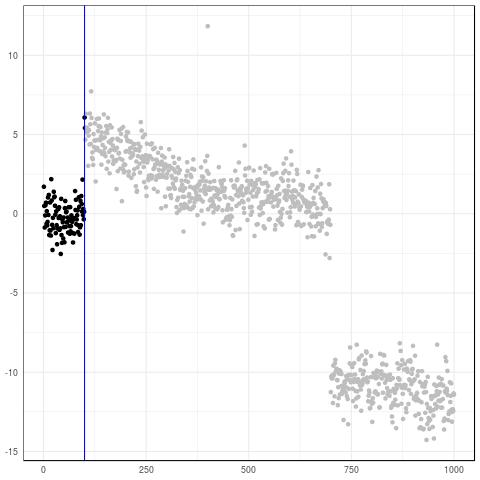}
		\caption{t=101}	\label{fig:rw_ex_101} 
	\end{subfigure}
	\begin{subfigure}[b]{0.32\linewidth}
	\centering
	\includegraphics[width=0.9\linewidth]{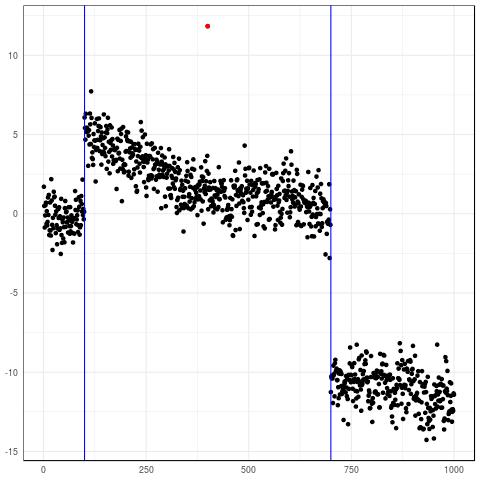}
	\caption{Full data}	\label{fig:rw_ex_1000} 
	\end{subfigure}
	\caption{Robust particle filter output at various times. Additive anomalies are denoted by red points, innovative anomalies by blue lines. Grey observations are yet to be observed.}
	\label{fig:rw_ex_solved}
\end{figure}

The proposed filter can be applied to the data displayed in Figure \ref{fig:rw_ex} to detect anomalies in an online fashion. It is worth pointing out that the filter re-evaluates past anomalies as more data becomes available. This can be seen in Figure \ref{fig:rw_ex_solved}:  When initially encountering the anomaly at time $t=100$ the filter gives approximately equal weight to the possibility of it being an additive outlier and to it being an innovative one. It is only when the next observation becomes available, that the filter (correctly) classifies it as an innovative anomaly. Note that only $N=20$ particles were used and only $M=1$ descendent of each anomaly type was sampled per particle.

\section{Particle Filter With Back-Sampling -- CE-BASS}\label{sec:Backsampling}

As mentioned in the introduction, it is possible that innovative outliers may not immediately be observed. One such example are innovative outliers in the trend component of the model described in \eqref{eq:RWand_T}. The filter as described in Algorithm \ref{alg:Basic} can not deal with such anomalies as it only inflates the variance of the innovative process at time $t$ when there is evidence in the observation at the same time $t$ that an outlier occurred. This can be remedied by back-sampling particles representing innovative outliers at a later time, $t+k$, once more observations and therefore evidence for an anomaly are available. This can be done using nearly identical approximation strategies as used in the previous section and allows to relax the assumptions made in the previous section from $\textbf{C}$ not having any $\textbf{0}$-columns to requiring that the system be observable. 

\subsection{Back-Sampling Particles Using the Last $k+1$ Observations}

The proposed back-sampling strategy at time $t$ consists of sampling particles for $(\textbf{V}_{t+1-k},...\textbf{V}_{t+1},\textbf{W}_{t+1-k},...,\textbf{W}_{t+1})$ given a $N(\bm{\mu}_{t-k},\bm{\Sigma}_{t-k} )$ filtering distribution for $\textbf{x}_{t-k}$ and observations $\textbf{Y}_{t-k+1},...,\textbf{Y}_{t-k}$. Specifically, we sample particles with a innovative single anomaly in $\textbf{W}_{t+1-k}$ assuming no other innovative anomalies or additive anomalies. Conditional on these augmented particles classical Kalman updates can once more be used as shown in Algorithm \ref{alg:Back-sample}. It should be noted that Algorithm \ref{alg:Basic} is a special case of Algorithm \ref{alg:Back-sample} which arises from setting $\mathcal{B}_1 = ... = \mathcal{B}_q = \{1\}$. 

\begin{algorithm}
	\caption{Particle Filter (With Back Sampling) -- CE-BASS}
	\label{alg:Back-sample}
	\begin{footnotesize}
		\begin{tabular}[h]{ll}
			{\bf Input:} & An initial state estimate $(\bm{\mu}_0,\bm{\Sigma}_0)$. \\ & A number of descendants, $M'=M(p+q)+1$. \\ & A number of particles to be maintained, $N$.
			\\ & A stream of observations $\textbf{Y}_1,\textbf{Y}_2,...$ \\ {\bf Initialise:} & Set $Particles(0) = \{(\bm{\mu}_0,\bm{\Sigma}_0,1)\}$ \\
			& Set $max\_horizon = \max \left(\cup_{i=1}^q \mathcal{B}_i\right)$
		\end{tabular}
		
		\begin{algorithmic}[1]
			\For{$t \in \mathbb{N}^+ $}
			\State $Cand \gets \{\}$ \Comment{To Store Candidates}
			\For{$(\bm{\mu},\bm{\Sigma},prob_{prev}) \in Particles(t-1)$}
			\State $(\textbf{V},\textbf{W},prob) \gets  \text{Sample\_typical}(\bm{\mu},\bm{\Sigma},\textbf{Y}_t,\textbf{A},\textbf{C},\bm{\Sigma}_A,\bm{\Sigma}_I)$
			\State $Cand \gets Cand \cup \{(\bm{\mu},\bm{\Sigma},\textbf{V},\textbf{W},prob\cdot prob_{prev},1)\}$
			\State $Add\_Des \gets \text{Sample\_additive}(\bm{\mu},\bm{\Sigma},\textbf{Y}_t,\textbf{A},\textbf{C},\bm{\Sigma}_A,\bm{\Sigma}_I,M)$
			\For {$(\textbf{V},\textbf{W},prob) \in Add\_Des$}
			\State $Cand \gets Cand \cup \{(\bm{\mu},\bm{\Sigma},\textbf{V},\textbf{W},prob\cdot prob_{prev},1)\}$
			\EndFor
			\EndFor
			\For {$hor \in \{1,...,max\_horizon\}$}
			\For{$(\bm{\mu},\bm{\Sigma},prob_{prev}) \in Particles(t-hor)$}
			\State $\tilde{\textbf{Y}} \gets \left[\textbf{Y}_{t-hor+1}^T,...,\textbf{Y}_{t}^T\right]^T$
			\State $Inn\_Des \gets \text{BS\_inn}(\bm{\mu},\bm{\Sigma},\tilde{\textbf{Y}},\textbf{A},\textbf{C},\bm{\Sigma}_A,\bm{\Sigma}_I,M,hor)$
			\For {$(\textbf{V},\textbf{W},prob) \in Inn\_Des$}
			\State $Cand \gets Cand \cup \{(\bm{\mu},\bm{\Sigma},\textbf{V},\textbf{W},prob\cdot prob_{prev},hor)\}$
			\EndFor 
			\EndFor
			\EndFor
			\State $Desc \gets \text{Subsample}(N,Cand)$
			\Comment{Sampling proportional to $prob$}
			\State $Particles(t) \gets \{\}$
			\For{$(\bm{\mu},\bm{\Sigma},\textbf{V},\textbf{W},prob,hor) \in Descendants $}
			\State $(\bm{\mu},\bm{\Sigma}) \gets  \text{KF\_Upd}(\textbf{Y}_{t+1-hor},\bm{\mu},\bm{\Sigma},\textbf{C},\textbf{A},\textbf{V}^{1/2}\bm{\Sigma}_A,\textbf{W}^{1/2}\bm{\Sigma}_I)$
			\If{$hor > 1$}
			\For {$i \in \{2,...,hor\}$}
			\State $(\bm{\mu},\bm{\Sigma}) \gets  \text{KF\_Upd}(\textbf{Y}_{t+i-hor},\bm{\mu},\bm{\Sigma},\textbf{C},\textbf{A},\bm{\Sigma}_A,\bm{\Sigma}_I)$
			\EndFor
			\EndIf
			\State $Particles(t) \gets Particles(t) \cup \{ (\bm{\mu},\bm{\Sigma},prob \cdot \frac{|Cand|}{|Desc|}) \}$
			\EndFor
			\EndFor
		\end{algorithmic}
	\end{footnotesize}
\end{algorithm}

To sample a particle with an innovative anomaly in the $j$th component of $\textbf{W}_{t+1-k}$, we define an augmented observation vector $\tilde{\textbf{Y}}_{t+1-k}^{(k)} = (\textbf{Y}_{t+1-k}^T,...,\textbf{Y}_{t+1}^T)^T$. This is normally distributed with mean $\tilde{\textbf{C}}^{(k)} \textbf{A}\bm{\mu}_{t-k}$ and variance
\footnotesize
\begin{equation*} 
\tilde{\textbf{C}}^{(k)} \left(  \textbf{A} \bm{\Sigma}_{t-k} \textbf{A}^T + \tilde{\textbf{Q}}^{(k)}  \right) \left(\tilde{\textbf{C}}^{(k)}\right)^T +
\tilde{\textbf{R}}^{(k)},
\end{equation*}
\normalsize
where $\tilde{\textbf{C}}^{(k)} = \textbf{C} \left(\left(\textbf{A}^0\right)^T,...,\left(\textbf{A}^k\right)^T\right)^T$ denotes the augmented matrix mapping the hidden states to the observations, 
\footnotesize
\begin{equation*} 
\tilde{\textbf{R}}^{(k)} = \begin{bmatrix}
\textbf{V}_{t+1-k}^{-1}  \bm{\Sigma}_A     & 0 & \ddots  \\
0      & \ddots & 0 \\
\ddots       & 0 & \textbf{V}_{t+1}^{-1}\bm{\Sigma}_A  
\end{bmatrix}
\end{equation*}
\normalsize
and
\footnotesize
\begin{equation*} 
\tilde{\textbf{Q}}^{(k)} = \begin{bmatrix}
\textbf{W}_{t+1-k}^{-1}  \bm{\Sigma}_I    & 0 & \ddots  \\
0      & \ddots & 0 \\
\ddots       & 0 & \textbf{W}_{t+1}^{-1}\bm{\Sigma}_I 
\end{bmatrix}
\end{equation*}
\normalsize
In a similar spirit, we define the augmented predictive variance to be
\begin{equation*}
\hat{\bm{\Sigma}}^{(k)} = \tilde{\textbf{C}}^{(k)} \left(  \textbf{A} \bm{\Sigma}_{t-k} \textbf{A}^T + \textbf{I}_{k+1} \otimes \bm{\Sigma}_I   \right) \left(\tilde{\textbf{C}}^{(k)}\right)^T + \textbf{I}_{k+1} \otimes
 \bm{\Sigma}_A  .
\end{equation*}
As a result of this reformulation, we retrieve update equations consisting of a single Kalman step, albeit with slightly different dimensions of the observation, $(k+1)p$ instead of $p$. It is therefore possible to use the sampling procedure for innovative outliers introduced in Section \ref{sec:Props}. This consists of sampling particles for $\tilde{\textbf{W}}_{t+1-k}^{(j,j)}$ from
\footnotesize
\begin{equation*}
\hat{\sigma_j}\Gamma\left(b_j + \frac{1}{2},b_j + \frac{\hat{\sigma_j}}{2\bm{\Sigma}_I^{(j,j)}}\left( \frac{\left(\left(\tilde{\textbf{C}}^{(k)}\right)^T\right)^{(j,:)}\left(\hat{\bm{\Sigma}}^{(k)}\right)^{-1} \tilde{\textbf{z}}_{t+1-k}^{(k)} }{\left(\left(\tilde{\textbf{C}}^{(k)}\right)^T\left(\hat{\bm{\Sigma}}^{(k)}\right)^{-1}\tilde{\textbf{C}}^{(k)}\right)^{(j,j)}}\right)^2\right).
\end{equation*}
\normalsize
for the residual $ \tilde{\textbf{z}}_{t+1-k}^{(k)} \tilde{\textbf{Y}}_{t+1-k}^{(k)}-\tilde{\textbf{C}}^{(k)} \textbf{A}\bm{\mu}_{t-k}$. The associated weight is given in Theorem 5 in the appendix. 

As in Section \ref{sec:weights}, we want to give different particles equal weights if they explain anomalies equally well. In particular, we therefore want to balance out the weights given to the back-sampled particles and the descendants of particles with an anomaly sampled at time $t-k+1$ using just $\textbf{Y}_{t+1-k}$. In order to do so, consider observations $\textbf{Y}_{t+1},...,\textbf{Y}_{t+1-k}$ which are such that they perfectly fit an innovative outlier in the $i$th innovative component at time $t-k+1$, i.e. 
\small
\begin{equation*}
\tilde{\textbf{Y}}_{t+1-k}^{(k)}-\left(\tilde{\textbf{C}}^{(k)}\right)\textbf{A}\bm{\mu}_{t-k} = \frac{\left(\tilde{\textbf{C}}^{(k)}\right)^{(:,j)}}{\sqrt{\left(\left(\tilde{\textbf{C}}^{(k)}\right)^T\left(\hat{\bm{\Sigma}}^{(k)}\right)^{-1}\left(\tilde{\textbf{C}}^{(k)}\right)\right)^{(j,j)}}}\delta.
\end{equation*}
\normalsize
As $\delta$ grows, the importance weight behaves as 
\begin{equation*}
\frac{b_j^{b_j}\frac{1}{M}s_j\frac{\Gamma(b_j+\frac{1}{2})}{\Gamma(b_j)} \exp\left(-\delta^2\right)
}{\left( 
	\frac{\hat{\sigma}_j}{2\bm{\Sigma}_I^{(j,j)} 
		\left(\left(\tilde{\textbf{C}}^{(k)}\right)^T\left(\hat{\bm{\Sigma}}^{(k)}\right)^{-1}\left(\tilde{\textbf{C}}^{(k)}\right)\right)^{(j,j)}
	}\delta^2
	\right)^{b_j}
},
\end{equation*}
up to the likelihood term and the $\left(1-\sum_{i=1}^{p}r_i -  \sum_{j=1}^{q}s_j\right)^{k}$ factor. However, these terms are also present in the weights of the descendants of the particles sampled at $t+1-k$ if no further anomaly was sampled at times $t+2-k,...,t+1$. Therefore, setting
\begin{equation*}
\hat{\sigma}_j = \bm{\Sigma}_I^{(j,j)} 
\left(\left(\tilde{\textbf{C}}^{(k)}\right)^T\left(\hat{\bm{\Sigma}}^{(k)}\right)^{-1}\left(\tilde{\textbf{C}}^{(k)}\right)\right)^{(j,j)}
\end{equation*}
results in the same asymptotic probabilities as the one obtained in Section \ref{sec:weights}. Given $\hat{\sigma}_j$ can only take a single value we set 
\begin{equation*}
\hat{\sigma}_j = \max_{k \in \mathcal{B}_j} \left( \bm{\Sigma}_I^{(j,j)} 
\left(\left(\tilde{\textbf{C}}^{(k)}\right)^T\left(\hat{\bm{\Sigma}}^{(k)}\right)^{-1}\left(\tilde{\textbf{C}}^{(k)}\right)\right)^{(j,j)} \right),
\end{equation*} 
where $\mathcal{B}_j \subset \mathbb{N}$ denotes the set of horizons used to back-sample the $j$th component of the $\textbf{W}_t$. 

A range of observations guide the choice of the sets $\mathcal{B}_j$ for $1 \leq j \leq q$. We assume that the Kalman model is observable, i.e.\ that there exists a $k$ such that the matrix $ \left[ \left(\textbf{C}\right)^T , \left(\textbf{CA}\right)^T, ... ,\left(\textbf{CA}^k\right)^T\right] $ has full column rank. Let $k^*$ denote the lowest such $k$. It is advisable to choose the set $\mathcal{B}_j$ such that it contains at least one element greater or equal to $k^*$. The reason for this being that any innovative anomaly capable of eventually influencing the observations must do so within $k^*$ observations from occurring. It should also be noted that a horizon $h$ can only be in the set $\mathcal{B}_j$ if the $j$th column of the augmented mapping from the hidden states to the observations, $\tilde{\textbf{C}}^{(h)}$, is non-zero as this is required by the proposal. Consequently, setting $\mathcal{B}_j = \left\{k \in \{1,...,k^*\} : \left(\tilde{\textbf{C}}^{(k)}\right)^{(:,j)} \neq \textbf{0}\right\}$ is a natural choice.

\subsection{Example}

\begin{figure}
	\begin{subfigure}[b]{0.32\linewidth}
		\centering
		\includegraphics[width=0.9\linewidth]{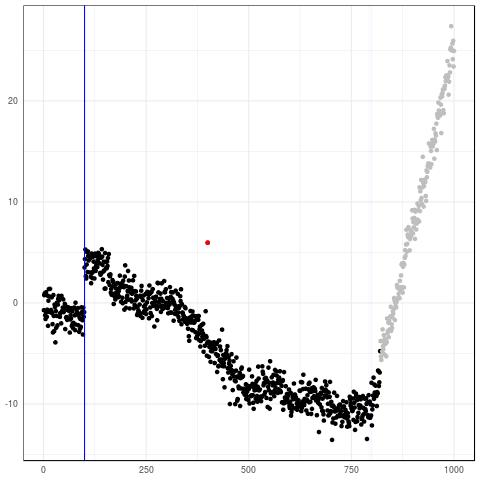}
		\caption{t=820} \label{fig:rwandt_ex_100} 
	\end{subfigure}
	\begin{subfigure}[b]{0.32\linewidth}
		\centering
		\includegraphics[width=0.9\linewidth]{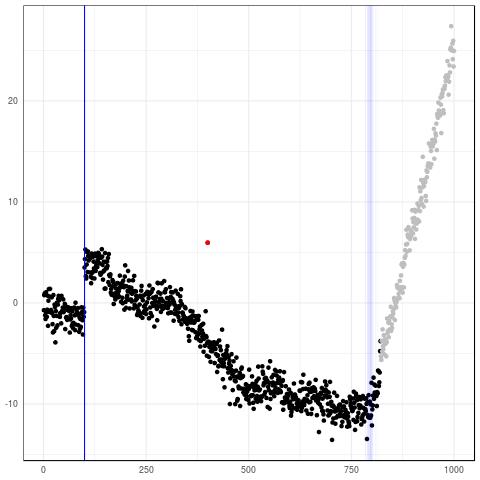}
		\caption{t=821}	\label{fig:rwandt_ex_101} 
	\end{subfigure}
	\begin{subfigure}[b]{0.32\linewidth}
		\centering
		\includegraphics[width=0.9\linewidth]{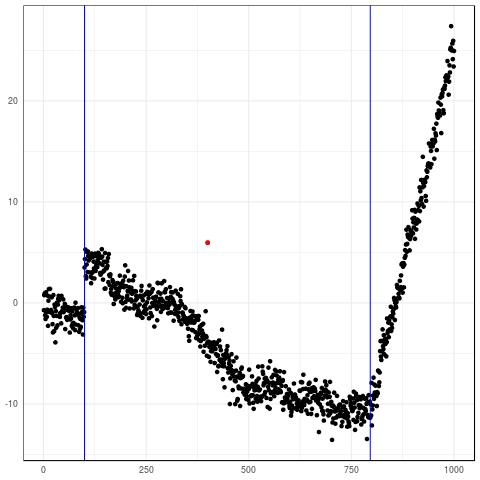}
		\caption{Full data}	\label{fig:rwandt_ex_1000} 
	\end{subfigure}
	\caption{Robust particle filter output at various times. Additive anomalies are denoted by red points, innovative anomalies by blue lines. Grey observations are yet to be observed.}
	\label{fig:rwandt_ex_solved}
\end{figure}

With back-sampling, we are now able to tackle the example from Figure \ref{fig:rwandt_ex}. We used $\mathcal{B}_1 = \{1,...,40\}$, $\mathcal{B}_2 = \{1,...,40\}$, to sample back up to 40 observations. We maintained $N=40$ particles and sampled $M=1$ descendants of each type. The output of the particle filter can be seen in Figure \ref{fig:rwandt_ex_solved}. As before, the filter updates its output as new observations become available. Whilst the trend innovation occurs at time $t=800$, the anomaly is first detected around time $t=820$. Even then, there is a large amount of uncertainty regarding the precise location of the anomaly which only gets resolved at a later time. 

\section{Simulations}\label{sec:Simulation}

\begin{figure} 
	\begin{subfigure}[b]{0.24\linewidth}
		\centering
		\includegraphics[width=0.95\linewidth]{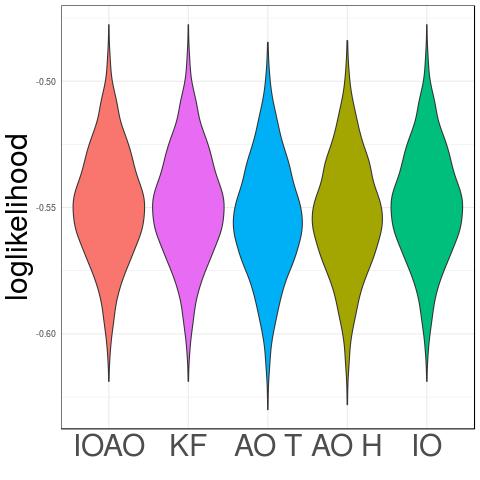} 
		\caption{Case 1}
		\label{fig:meanchange_graph} 
	\end{subfigure} 
	\begin{subfigure}[b]{0.24\linewidth}
		\centering
		\includegraphics[width=0.95\linewidth]{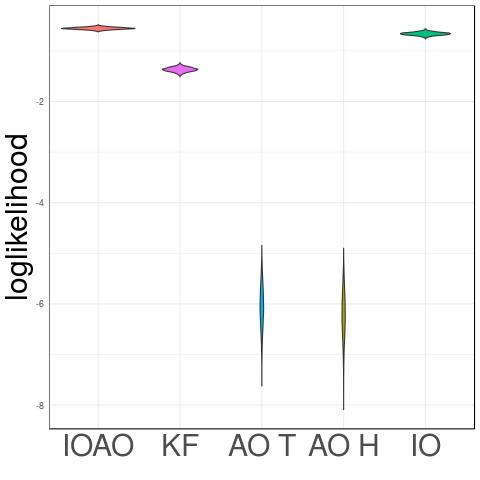} 
		\caption{Case 1, IOs} 
		\label{fig:meanANOMchange_graph} 
	\end{subfigure} 
		\begin{subfigure}[b]{0.24\linewidth}
		\centering
		\includegraphics[width=0.95\linewidth]{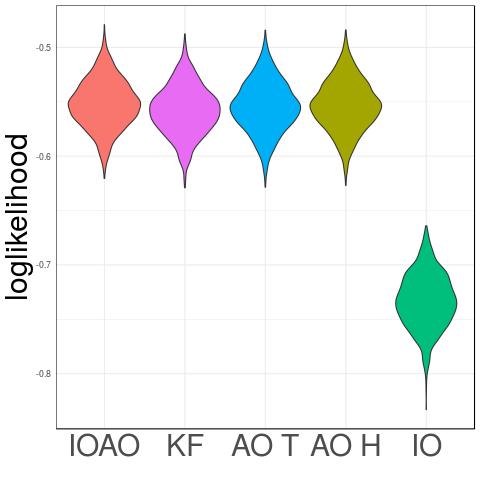} 
		\caption{Case 1, AOs} 
		\label{fig:meanANOMchange_graph} 
	\end{subfigure} 
	\begin{subfigure}[b]{0.24\linewidth}
		\centering
		\includegraphics[width=0.95\linewidth]{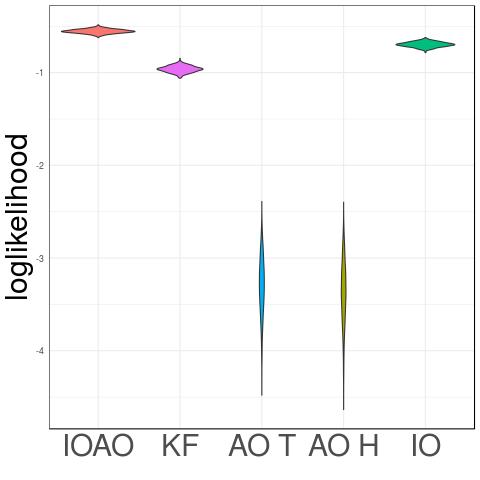}
		\caption{Case 1, Both}
		\label{fig:meanchange_graph} 
	\end{subfigure}
		\begin{subfigure}[b]{0.24\linewidth}
		\centering
		\includegraphics[width=0.95\linewidth]{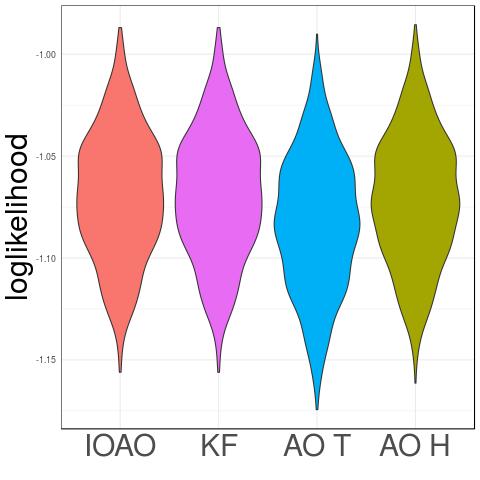} 
		\caption{Case 2}
		\label{fig:meanchange_graph} 
	\end{subfigure} 
	\begin{subfigure}[b]{0.24\linewidth}
		\centering
		\includegraphics[width=0.95\linewidth]{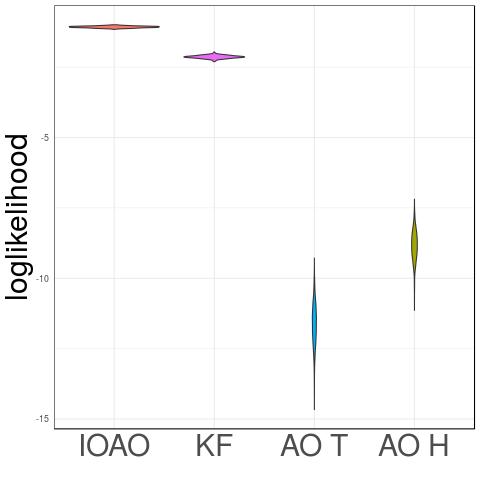} 
		\caption{Case 2, IOs} 
		\label{fig:meanANOMchange_graph} 
	\end{subfigure} 
		\begin{subfigure}[b]{0.24\linewidth}
		\centering
		\includegraphics[width=0.95\linewidth]{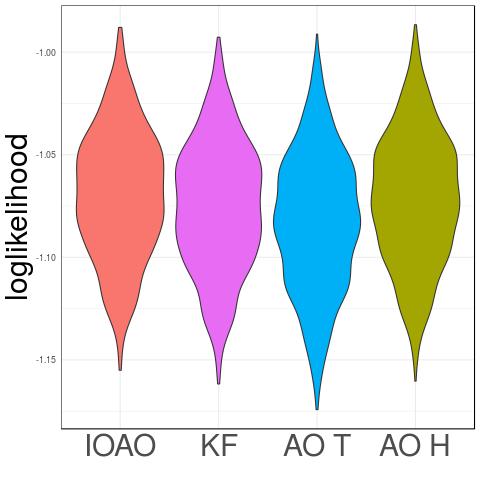} 
		\caption{Case 2, AOs} 
		\label{fig:meanANOMchange_graph} 
	\end{subfigure} 
	\begin{subfigure}[b]{0.24\linewidth}
		\centering
		\includegraphics[width=0.95\linewidth]{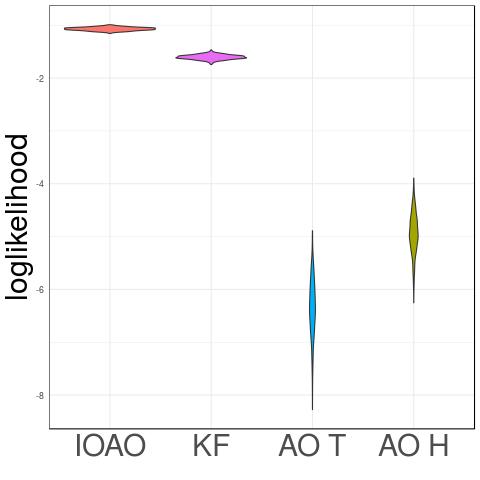}
		\caption{Case 2, Both}
		\label{fig:meanchange_graph} 
	\end{subfigure}
			\begin{subfigure}[b]{0.24\linewidth}
		\centering
		\includegraphics[width=0.95\linewidth]{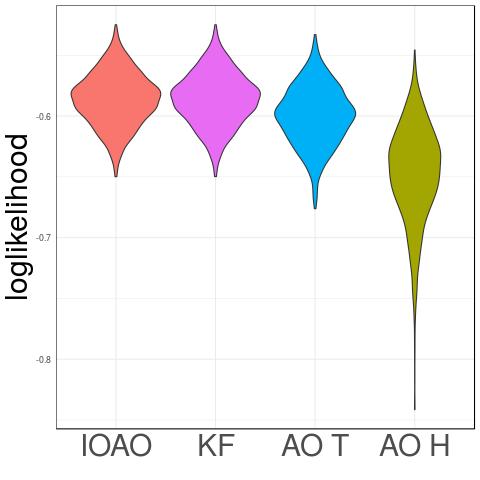} 
		\caption{Case 3}
		\label{fig:meanchange_graph} 
	\end{subfigure} 
	\begin{subfigure}[b]{0.24\linewidth}
		\centering
		\includegraphics[width=0.95\linewidth]{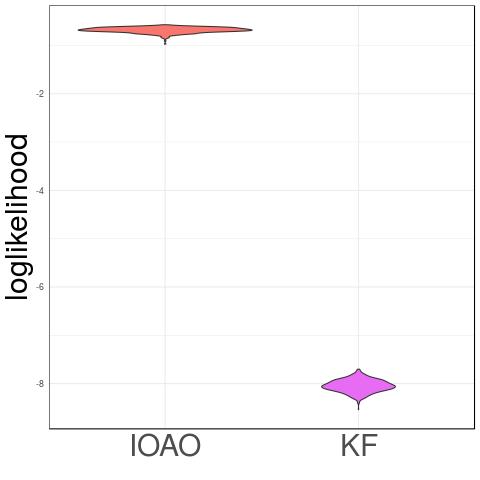} 
		\caption{Case 3, IOs} 
		\label{fig:meanANOMchange_graph} 
	\end{subfigure} 
		\begin{subfigure}[b]{0.24\linewidth}
		\centering
		\includegraphics[width=0.95\linewidth]{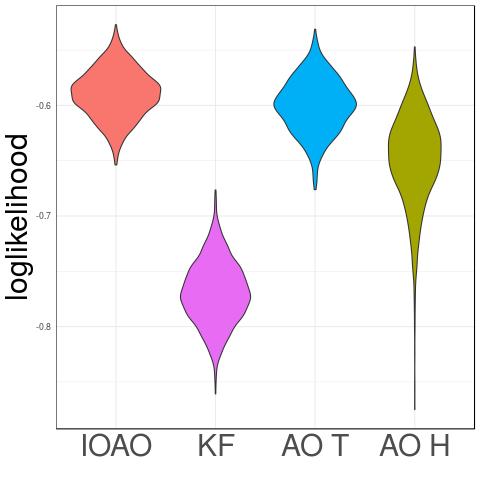} 
		\caption{Case 3, AOs} 
		\label{fig:meanANOMchange_graph} 
	\end{subfigure} 
	\begin{subfigure}[b]{0.24\linewidth}
		\centering
		\includegraphics[width=0.95\linewidth]{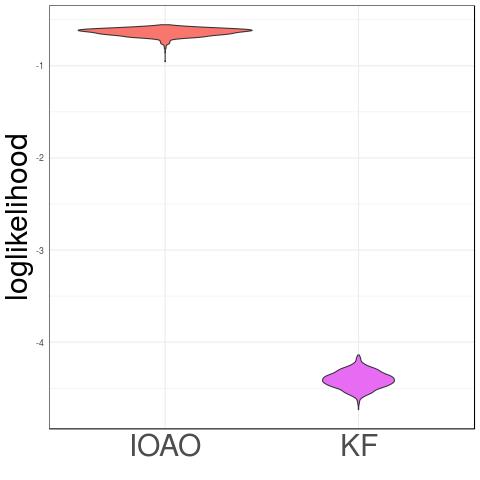}
		\caption{Case 3, Both}
		\label{fig:meanchange_graph} 
	\end{subfigure}
			\begin{subfigure}[b]{0.24\linewidth}
		\centering
		\includegraphics[width=0.95\linewidth]{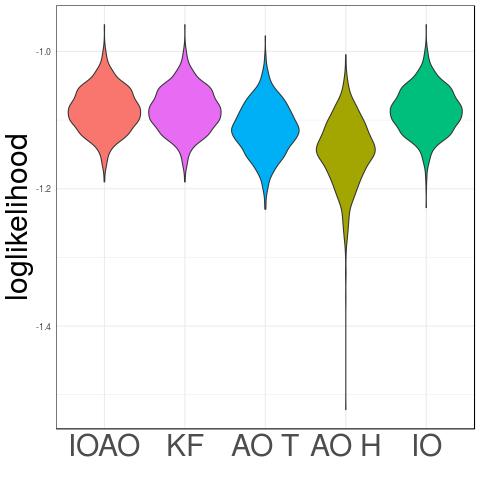} 
		\caption{Case 4}
		\label{fig:meanchange_graph} 
	\end{subfigure} 
	\begin{subfigure}[b]{0.245\linewidth}
		\centering
		\includegraphics[width=0.95\linewidth]{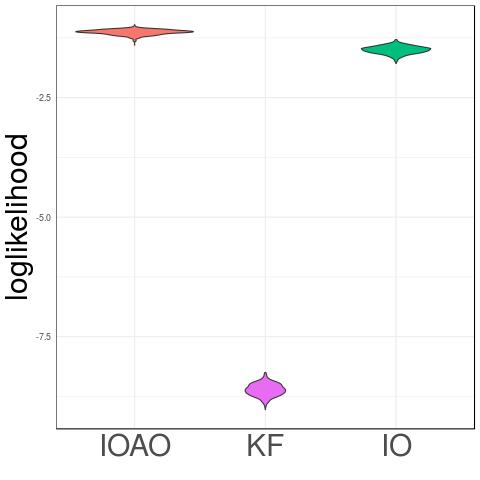} 
		\caption{Case 4, IOs} 
		\label{fig:meanANOMchange_graph} 
	\end{subfigure} 
		\begin{subfigure}[b]{0.24\linewidth}
		\centering
		\includegraphics[width=0.95\linewidth]{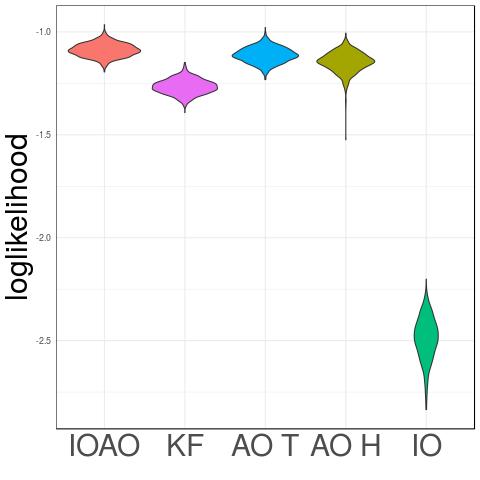} 
		\caption{Case 4, AOs} 
		\label{fig:meanANOMchange_graph} 
	\end{subfigure} 
	\begin{subfigure}[b]{0.24\linewidth}
		\centering
		\includegraphics[width=0.95\linewidth]{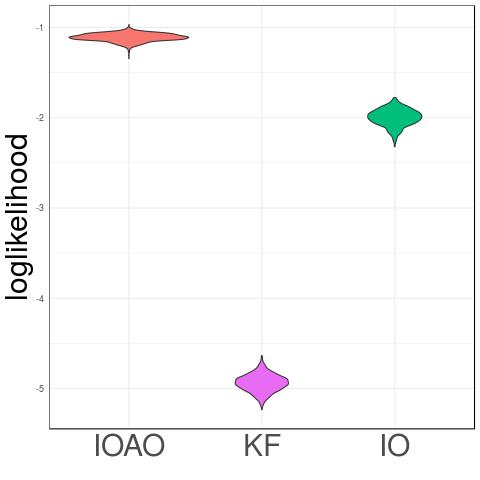}
		\caption{Case 4, Both}
		\label{fig:meanchange_graph} 
	\end{subfigure}
	\caption{Violin plots for the average predictive log-likelihood of the five filters (IOAO: CE-BASS, KF: The classical Kalman Filter, AO T: \cite{agamennoni2011outlier}, AO H: \cite{ruckdeschel2014robust}, IO H: \cite{ruckdeschel2014robust}) over the four different scenarios under a range of models. Higher values correspond to better performance. Methods are omitted on the graphs if they can not be applied to the setting or if their performance is too poor.}
	\label{fig:LOG-LIK} 
\end{figure}

We now turn to comparing CE-BASS against other methods. In particular, we compare against the $t$-distribution based additive outlier robust filter by \cite{agamennoni2011outlier}, the Huberisation based additive outlier robust filter by \cite{ruckdeschel2014robust}, the Huberisation based innovative outlier robust filter by \cite{ruckdeschel2014robust}, and the classical Kalman Filter \citep{Kalman1960}. All these algorithms are implemented in the accompanying package. 

We consider four different models and generate 1000 observations for each. For each of the four models, we consider a case in which no anomalies are present, a case in which only additive anomalies are present, a case in which only innovative anomalies are present, and a case in which both additive and innovative anomalies are present. When anomalies are added, they are added at times $t=100$, $t=300$, $t=600$, and $t=900$. Specifically we considered the following three models:
\begin{enumerate}
    \item The model of Example 1 with $\sigma_A=1$ and $\sigma_I=0.1$.
 We consider a case with only additive outliers, a case with only innovative outliers, and a case where an additive outlier at $t=100$, is followed by two innovative outliers at times $t=300$ and $t=600$, which were then followed by an additive outlier at time $t=900$. To simulate additive anomalies, we set $V_t^{\frac{1}{2}}\sigma_A\epsilon_t = 10$ and to simulate the innovative outliers we set $W_t^{\frac{1}{2}}\sigma_I\nu_t = 10$.
    \item The random walk model with two measurements
    \small
    \begin{align*}
Y_t^{(1)} &= X_t + \left( V_t^{(1)} \right)^{\frac{1}{2}}\sigma_A^{(1)}\epsilon_t^{(1)}, \;\; &  \;\; X_t = X_{t-1} + W_t^{\frac{1}{2}}\sigma_I\nu_t \\
Y_t^{(2)} &= X_t + \left( V_t^{(2)} \right)^{\frac{1}{2}}\sigma_A^{(2)}\epsilon_t^{(2)}, \;\; & 
    \end{align*}
    \normalsize
    where $\sigma_A^{(1)} = \sigma_A^{(2)} = 1$ for $i=1,2$ and $\sigma_I=0.1$. We consider a case with only additive outliers (one in the first component, then two in the second, then one in the first), a case with only innovative outliers, and a case where an additive outlier in the first component at time $t=100$ is followed by two innovative outliers at times $t=300$ and $t=600$, which are then followed by an additive outlier in the second component at time $t=900$. For additive anomalies, we set $\left( V_t^{(1)} \right)^{\frac{1}{2}}\sigma_A^{(1)}\epsilon_t^{(1)} = 10$ or $\left( V_t^{(2)} \right)^{\frac{1}{2}}\sigma_A^{(2)}\epsilon_t^{(2)} = 10$ and for innovative outliers, we set $W_t^{\frac{1}{2}}\sigma_I\nu_t = 10$.
    \item 
The model of Example 2 with $\sigma_A=1$, $\sigma_I^{(1)}=0.1$ and $\sigma_I^{(2)}=0.01$. 
We consider a case with only additive outliers, a case with only innovative outliers (one in the second component, then one in the first, then one in the second, then one in the first), and a case with an additive outlier at $t=100$, followed by an innovative outlier affecting the first component of the hidden state at times $t=300$, followed by an innovative outlier affecting the second component of the hidden state at times $t=600$, followed by an additive outlier at time $t=900$. The additive anomalies were instances where we set $V_t^{\frac{1}{2}}\epsilon_t = 30$  and the innovative outliers were instances where we set $\left( W_t^{(1)} \right)^{\frac{1}{2}}\eta_t^{(1)} = 100$ or $\left( W_t^{(2)} \right)^{\frac{1}{2}}\eta_t^{(2)} = 500$.
    \item 
An extension of Example 2 where the position is also observed. The equations governing the hidden state are as before whilst the equations governing the observations are    
        \footnotesize
    \begin{align*}
Y_t^{(1)} &= X_t^{(1)} + \left( V_t^{(1)} \right)^{\frac{1}{2}}\sigma_A^{(1)}\epsilon_t^{(1)},  \\
Y_t^{(2)} &= X_t^{(2)} + \left( V_t^{(2)} \right)^{\frac{1}{2}}\sigma_A^{(2)}\epsilon_t^{(2)},
    \end{align*}
    \normalsize
where $\sigma_A^{(1)} = \sigma_A^{(2)} = 1$. We consider a case with only additive outliers (in the first component only), a case with only innovative outliers (one in the second component, then one in the first, then one in the second, then one in the first), and a case with an additive outlier at time $t=100$, followed by an innovative outlier affecting the first component of the hidden state at time $t=300$, followed by an innovative outlier affecting the second component of the hidden state at time $t=600$, followed by an additive outlier at time $t=900$. For additive anomalies, we set $\left( V_t^{(1)} \right)^{\frac{1}{2}}\sigma_A^{(1)}\epsilon_t^{(1)} = 30$ and for innovative outliers, we set $\left( W_t^{(1)} \right)^{\frac{1}{2}}\sigma_I^{(1)}\eta_t^{(1)} = 100$ or $\left( W_t^{(2)} \right)^{\frac{1}{2}}\sigma_I^{(2)}\eta_t^{(2)} = 500$.
\end{enumerate}

We evaluate the different methods based on average predictive log-likelihood and average predictive mean squared error. We exclude all observations corresponding to anomalies from the calculation of these averages since the filters can not be expected to predict them. When calculating the average mean squared error we additionally remove one observation after the anomaly in the first setting and two observations in the third setting from the performance metric. This is to give the filter enough information to determine which type of anomaly the outlier corresponds to and return to a unimodal posterior, as the MSE is only an appropriate metric for unimodal posteriors.

The average log-likelihoods across all models can be found in Figure \ref{fig:LOG-LIK}, while the qualitatively very  similar results for the mean squared error can be found in the appendix. We see that the performance of CE-BASS compares favourably with that of the competing methods. In particular it is as accurate as the Kalman filter in the absence of anomalies and is more accurate than the additive outlier and innovative outlier robust filters even when only additive or innovative outliers are present, i.e.\ the settings for which these algorithms were designed. 

\section{Application}\label{sec:Application}

In this section, we apply CE-BASS to two real datasets.  We will use different types of models for the two applications to illustrate the way in which CE-BASS can be used. The first dataset is a labelled benchmark dataset which consists of temperature readings on a large industrial machine. Here, we will use a model which considerably restricts the movements of the hidden states when no anomalies are present, and thus emulates a changepoint model. The second is an unlabelled dataset which consist of repeated throughput measurements on a router. For that application we will use a model which has a considerable amount of flexibility and where the hidden states tend to follow the observations and therefore detect localised anomalies.

	\subsection{Machine Temperature Data}
	
\begin{figure} 
\begin{subfigure}[b]{0.495\linewidth}
	\centering
	\includegraphics[width=0.95\linewidth]{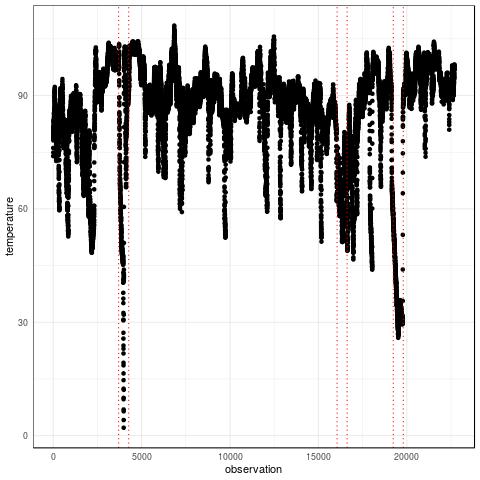} 
	\caption{Raw data with labels} 
	\label{fig:Rawdata} 
	\vspace{20pt} 
\end{subfigure} 
\begin{subfigure}[b]{0.495\linewidth}
	\centering
	\includegraphics[width=0.95\linewidth]{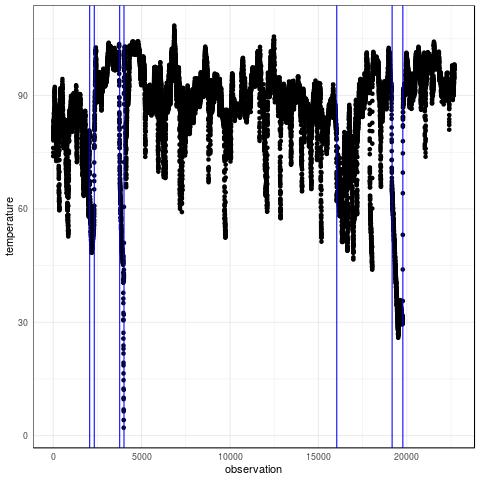} 
	\caption{CE-BASS output} 
	\label{fig:Analyseddata} 
	\vspace{20pt} 
\end{subfigure} 
	\caption{Machine temperature dataset. The labelled anomalies are: a planned shutdown, an early warning sign of a problem, and the catastrophic system failure caused by the problem.}
	\label{fig:Machine_Temp} 
\end{figure}
	
	We now apply CE-BASS to the machine temperature data taken from the Numenta Anomaly Benchmark (NAB, \cite{lavin2015evaluating}) which can be accessed at \textit{https://github.com/numenta/NAB}. The data consists of over 20000 readings from a temperature sensor on a large industrial machine and is displayed in Figure \ref{fig:Rawdata} along the three periods of anomalous behaviour labelled by an engineer. The first corresponds to a planned shutdown and the second to an early warning sign of the third anomaly -- a catastrophic failure. 
	
	In order to do so, we use the random walk model from Example 1 with the aim of detecting persistent changes in mean. We therefore use a maximum backsampling horizon of 250 by setting $\mathcal{B}_1=\{1,5,10,20,40,80,150,250\}$ and fix $\sigma_I = 1/10000\sigma_A$ to ensure that long and weak anomalies will not be interpreted as a persistent shift in the typical state. We use the first 15\% of the data, marked by \cite{lavin2015evaluating} as train data, to estimate the standard deviation $\sigma_A$ as well as the initial mean $\mu_0$ using the median absolute deviation and the median respectively. Using robust covariance methods we also detect very strong auto-correlation ($\rho=0.99$) and therefore took the default probabilities for anomalies to the power of $\frac{1}{1-\rho}$.
	
	The results of this analysis can be seen in Figure \ref{fig:Analyseddata}. We note that all anomalies flagged by the engineer are also being detected by CE-BASS. Two additional innovative anomalies around a prolonged drop which preceded the planned shutdown are also detected. They could be a false positive or an early warning sign of an anomaly prevented by the shutdown which has not been noticed by the engineer. 

	\subsection{Router Data}
	
		\begin{figure} 
			\begin{subfigure}[b]{0.32\linewidth}
				\centering
				\includegraphics[width=0.95\linewidth]{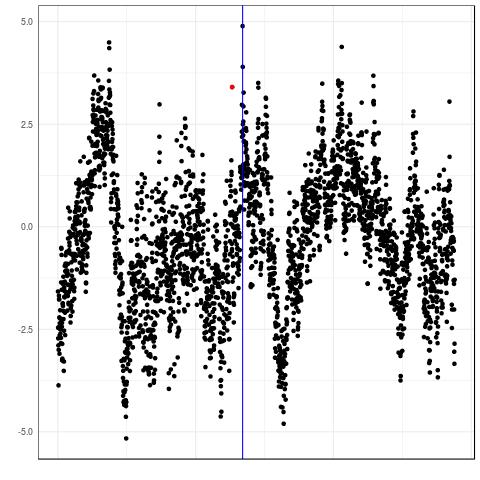} 
				\caption{Day 11}
				\label{fig:Day11} 
				\vspace{20pt} 
			\end{subfigure} 
			\begin{subfigure}[b]{0.32\linewidth}
				\centering
				\includegraphics[width=0.95\linewidth]{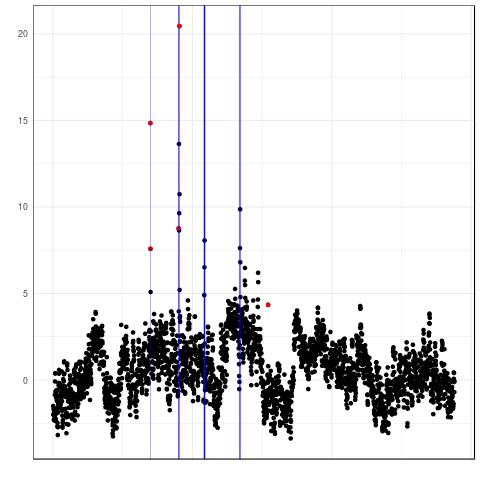} 
				\caption{Day 12} 
				\label{fig:Day12} 
				\vspace{20pt} 
			\end{subfigure} 
			\begin{subfigure}[b]{0.32\linewidth}
				\centering
				\includegraphics[width=0.95\linewidth]{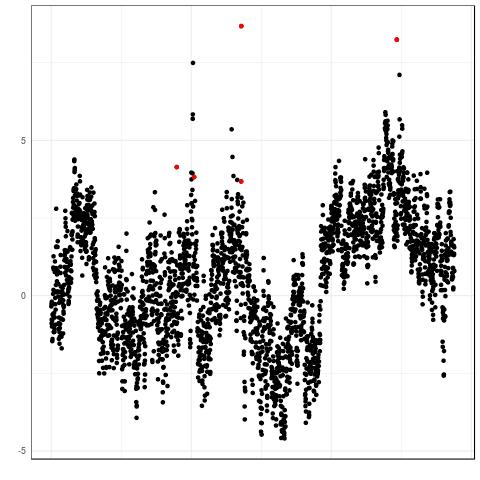} 
				\caption{Day 13} 
				\label{fig:Day13} 
				\vspace{20pt} 
			\end{subfigure}
			\begin{subfigure}[b]{0.32\linewidth}
	\centering
	\includegraphics[width=0.95\linewidth]{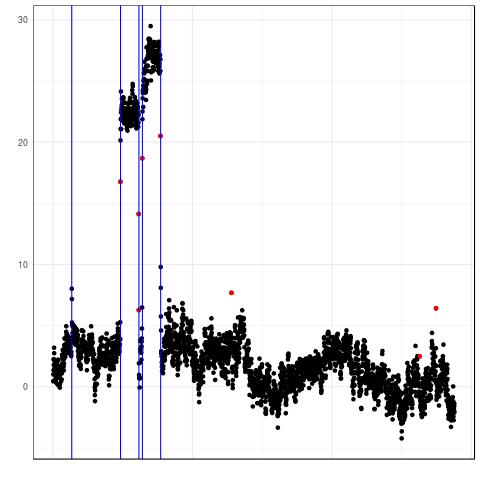} 
	\caption{Day 14}
	\label{fig:Day14} 
	\vspace{20pt} 
\end{subfigure} 
\begin{subfigure}[b]{0.32\linewidth}
	\centering
	\includegraphics[width=0.95\linewidth]{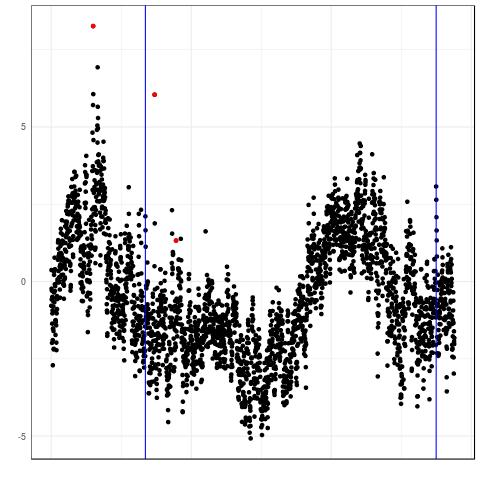} 
	\caption{Day 15} 
	\label{fig:Day15} 
	\vspace{20pt} 
\end{subfigure} 
\begin{subfigure}[b]{0.32\linewidth}
	\centering
	\includegraphics[width=0.95\linewidth]{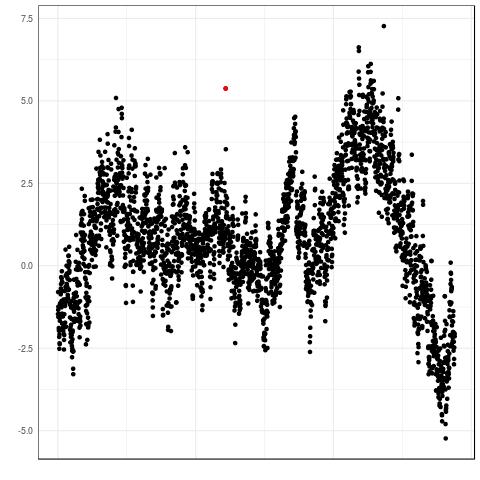} 
	\caption{Day 16} 
	\label{fig:Day16}
	\vspace{20pt}  
\end{subfigure} 
			\begin{subfigure}[b]{0.32\linewidth}
	\centering
	\includegraphics[width=0.95\linewidth]{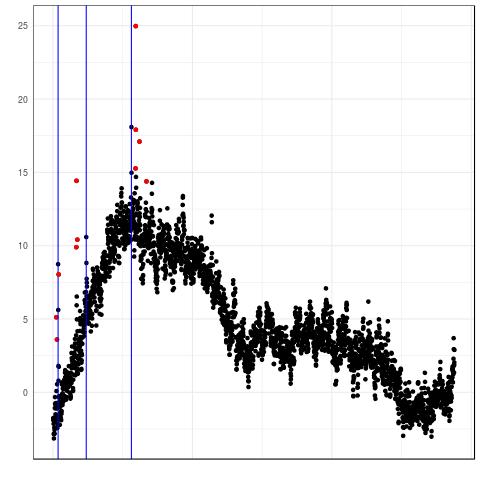} 
	\caption{Day 17}
	\label{fig:Day17}
	\vspace{20pt}  
\end{subfigure} 
\begin{subfigure}[b]{0.32\linewidth}
	\centering
	\includegraphics[width=0.95\linewidth]{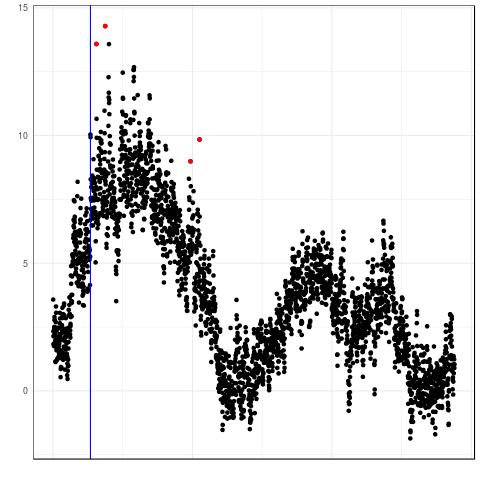} 
	\caption{Day 18} 
	\label{fig:Day18} 
	\vspace{20pt} 
\end{subfigure} 
\begin{subfigure}[b]{0.32\linewidth}
	\centering
	\includegraphics[width=0.95\linewidth]{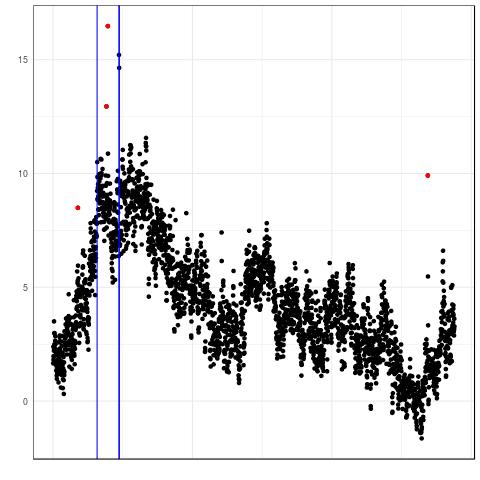} 
	\caption{Day 19} 
	\label{fig:Day19} 
	\vspace{20pt} 
\end{subfigure} 
			\caption{CE-BASS applied to 9 days of de-seasonalised router data. Lines correspond to innovative anomalies, i.e.\ spikes or level shifts.}
			\label{fig:Router} 
		\end{figure}
	
	The online analysis of aggregated traffic data on servers is an important challenge in both predictive maintenance and cyber security. This is because anomalies in throughput can point towards problems in the network such as malfunctions or malicious behaviour. Detecting anomalies as soon as possible therefore means that the root cause can be addressed more quickly -- potentially even before user experience is affected or harm caused.
	
	In this section, we consider 19 days worth of data from a network IP router which has been gathered at a frequency of one observation every 30 seconds. To preserve confidentiality, we de-seasonalised the data for days 11 to 19 using a seasonality model trained on days 1 to 10 and, for the purpose of this paper, consider only the de-seasonalised data for days 11 to 19 which can be found in Figures \ref{fig:Day11} to \ref{fig:Day19}. The main features apparent in the daily series are spikes, outliers, and changepoints. In order to capture these, we use an AR(1) model with slowly changing mean to model the observations $Y_t$. Formally, we used the model 
	\begin{align*}
	 	Y_t &= X_t^{(1)} + X_t^{(2)} + V_t\sigma_A \epsilon_t, \;\; & \;\;   X_t^{(1)} &= X_{t-1}^{(1)} + W_t^{(1)}\sigma_I^{(1)} \eta_t^{(1)}, \\
	 	& \;\;& \;\; X_t^{(2)} &= \rho X_{t-1}^{(2)} + W_t^{(2)}\sigma_I^{(2)} \eta_t^{(2)}.   
	\end{align*}
	Here, anomalies in $\epsilon_t$ correspond to isolated outliers, anomalies in $\eta_t^{(1)}$ correspond to level shifts and outliers in $\eta_t^{(2)}$ correspond to spikes. 
	
	We use the first 1000 observations of the first day, to obtain the estimates $\sigma_A = 0.0516$, $\sigma_I^{(1)} = 0.0157$, $\sigma_I^{(2)} = 0.516$, and $\rho = 0.815$. The result obtained from running CE-BASS with these parameters on the daily router data is displayed in Figures \ref{fig:Day11} to \ref{fig:Day19}. We note that very few of the anomalies returned can be classed as false positives. At the same time, a large number of anomalies are flagged, including a large number of outliers and spikes, but also some level shifts (Day 14). Discussion with engineers highlighted that the anomalies detected matched well with their knowledge of the data. This shows CE-BASS's ability to return a large number of diverse features which can be used as inputs to a supervised algorithm should labels become available.

\section{Acknowledgements} 

This work was supported by EPSRC grant numbers EP/N031938/1 (StatScale) and EP/L015692/1 (STOR-i). The authors also acknowledge British Telecommunications plc (BT) for financial support, David Yearling and Trevor Burbridge in BT Research for discussions.

\bibliographystyle{unsrt}  
\bibliography{references}

\newpage

\section{Appendix}

\setcounter{Thm}{0}

\subsection{Theorems and Derivations}\label{sec:Thms}

\subsubsection{Theorem \ref{Thm:V_Anom}}\label{sec:ThmV_Anom}

\begin{Thm}\label{Thm:V_Anom}
	Let the prior for the hidden state $\textbf{X}_{t}$ be $N(\bm{\mu},\bm{\Sigma})$ and an observation $\bm{Y}_{t+1} := \bm{Y}$ be available. Then the samples for $\tilde{\textbf{V}}^{(i,i)}_{t+1}$ from
	\begin{equation*}
	\tilde{\sigma}_i\Gamma\left(a_i + \frac{1}{2},a_i + \frac{\tilde{\sigma}_i}{2\bm{\Sigma}_A^{(i,i)}}\left( \frac{\left(\hat{\bm{\Sigma}}^{-1}\right)^{(i,:)} \left(\textbf{Y}-\textbf{C}\textbf{A}\bm{\mu}\right) }{\left(\hat{\bm{\Sigma}}^{-1}\right)^{(i,i)}}\right)^2\right)
	\end{equation*}
	have associated weight
	\footnotesize
	\begin{align*}
	\frac{1}{M}r_i\frac{\Gamma(a_i+\frac{1}{2})}{\Gamma(a_i)} \sqrt{\tilde{\sigma}_i} \frac{a_i^{a_i}}{\left( 
		a_i + \frac{\tilde{\sigma}_i}{2\bm{\Sigma}_A^{(i,i)}}\left( \frac{\left(\hat{\bm{\Sigma}}^{-1}\right)^{(i,:)} \left(\textbf{Y}-\textbf{C}\textbf{A}\bm{\mu}\right) }{\left(\hat{\bm{\Sigma}}^{-1}\right)^{(i,i)}}\right)^2	
		\right)^{a_i + \frac{1}{2}}} 
	\frac{
		\exp \left( 
		-\frac{1}{2} \left(\textbf{Y}-\textbf{C}\textbf{A}\bm{\mu}\right)^T
		\hat{\bm{\Sigma}}^{-1} 
		\left(\textbf{Y}-\textbf{C}\textbf{A}\bm{\mu}\right)
		\right)
	}{\sqrt{|\hat{\bm{\Sigma}}|} \sqrt{\left( \tilde{\textbf{V}}^{(i,i)} + \bm{\Sigma}_A^{(i,i)} \left(\hat{\bm{\Sigma}}^{-1}\right)^{(i,i)}\right)}  } \\
		\exp \left( 
		\frac{1}{2} \left( 1 +\left(\frac{\tilde{\textbf{V}}_{t+1}^{(i,i)}}{\bm{\Sigma}_A^{(i,i)} \left(\hat{\bm{\Sigma}}^{-1}\right)^{(i,i)} }\right)^2 \frac{\bm{\Sigma}_A^{(i,i)} \left(\hat{\bm{\Sigma}}^{-1}\right)^{(i,i)}}{\bm{\Sigma}_A^{(i,i)} \left(\hat{\bm{\Sigma}}^{-1}\right)^{(i,i)} +  \tilde{\textbf{V}}_{t+1}^{(i,i)}} \right)
		\left( \frac{\left(\hat{\bm{\Sigma}}^{-1}\right)^{(i,:)} \left(\textbf{Y}-\textbf{C}\textbf{A}\bm{\mu}\right) }{\sqrt{\left(\hat{\bm{\Sigma}}^{-1}\right)^{(i,i)}}}\right)^2
		\right).
	\end{align*}
	\normalsize
\end{Thm}

\textbf{Proof}: We wish to sample from the posterior distribution of $\tilde{\textbf{V}}^{(i,i)}_{t+1}$ which is proportional to 
\footnotesize
\begin{equation}\label{eq:MAIN}
r_i f_i\left(\tilde{\textbf{V}}_{t+1}^{(i,i)}\right)
\frac{
	\exp \left( 
	-\frac{1}{2} \left(\textbf{Y}-\textbf{C}\textbf{A}\bm{\mu}\right)^T
	\left( \hat{\bm{\Sigma}} +  \frac{\bm{\Sigma}_A^{(i,i)}}{\tilde{\textbf{V}}_{t+1}^{(i,i)}} \textbf{I}^{(i)}\right)^{-1} 
	\left(\textbf{Y}-\textbf{C}\textbf{A}\bm{\mu}\right)
	\right)
}{\sqrt{\left|\hat{\bm{\Sigma}} +  \frac{\bm{\Sigma}_A^{(i,i)}}{\tilde{\textbf{V}}_{t+1}^{(i,i)}} \textbf{I}^{(i)} \right|}},
\end{equation} \normalsize
where $f_i()$ denotes the PDF of a $\tilde{\sigma}_i\Gamma(a_i,a_i)$-distribution. The intractable part in the above consists of
\begin{equation*}
\left(\hat{\bm{\Sigma}} +  \frac{\bm{\Sigma}_A^{(i,i)}}{\tilde{\textbf{V}}_{t+1}^{(i,i)}} \textbf{I}^{(i)}\right)^{-1}, 
\end{equation*}
where $\textbf{I}^{(i)} = \textbf{e}_i \textbf{e}_i^T$ is a matrix which is 0 everywhere with the exception of the $i$th entry of the $i$th row, which is 1. Note that $ \textbf{I}^{(i)}$ has rank 1 and therefore, by the Sherman Morrison formula,
\begin{align*}
\left(\hat{\bm{\Sigma}} +  \frac{\bm{\Sigma}_A^{(i,i)}}{\tilde{\textbf{V}}_{t+1}^{(i,i)}} \textbf{I}^{(i)}\right)^{-1}
=
\hat{\bm{\Sigma}}^{-1} - \frac{\hat{\bm{\Sigma}}^{-1} \textbf{I}^{(i)}  \hat{\bm{\Sigma}}^{-1}}{1+tr(\hat{\bm{\Sigma}}^{-1}\textbf{I}^{(i)}) \frac{\bm{\Sigma}_A^{(i,i)}}{\tilde{\textbf{V}}_{t+1}^{(i,i)}} }\frac{\bm{\Sigma}_A^{(i,i)}}{\tilde{\textbf{V}}_{t+1}^{(i,i)}}  =  \hat{\bm{\Sigma}}^{-1} - \frac{1}{tr(\hat{\bm{\Sigma}}^{-1}\textbf{I}^{(i)})} \frac{\hat{\bm{\Sigma}}^{-1} \textbf{I}^{(i)}  \hat{\bm{\Sigma}}^{-1}}{1+ \frac{1}{tr(\hat{\bm{\Sigma}}^{-1}\textbf{I}^{(i)})\bm{\Sigma}_A^{(i,i)}} \tilde{\textbf{V}}_{t+1}^{(i,i)}}.
\end{align*}
Furthermore, given $tr(\hat{\bm{\Sigma}}^{-1}\textbf{I}^{(i)}) = \left(\hat{\bm{\Sigma}}^{-1}\right)^{(i,i)}$, the above is equal to
\begin{align*}
\hat{\bm{\Sigma}}^{-1} - \hat{\bm{\Sigma}}^{-1} \textbf{I}^{(i)}  \hat{\bm{\Sigma}}^{-1} \left[\frac{1}{\left(\hat{\bm{\Sigma}}^{-1}\right)^{(i,i)}} -  \left(\frac{1}{\left(\hat{\bm{\Sigma}}^{-1}\right)^{(i,i)}}\right)^2\frac{\tilde{\textbf{V}}_{t+1}^{(i,i)}}{\bm{\Sigma}_A^{(i,i)}}  + \left(\frac{\tilde{\textbf{V}}_{t+1}^{(i,i)}}{\bm{\Sigma}_A^{(i,i)} \left(\hat{\bm{\Sigma}}^{-1}\right)^{(i,i)} }\right)^2 \frac{1}{\left(\hat{\bm{\Sigma}}^{-1}\right)^{(i,i)} + \frac{1}{\bm{\Sigma}_A^{(i,i)}} \tilde{\textbf{V}}_{t+1}^{(i,i)}}\right].
\end{align*}
Crucially, the first term is constant in $\tilde{\textbf{V}}_{t+1}^{(i,i)}$, while the second is linear in $\tilde{\textbf{V}}_{t+1}^{(i,i)}$ and therefore conjugate to the prior of $\tilde{\textbf{V}}^{(i,i)}_{t+1}$. The last term is quadratic in $\tilde{\textbf{V}}^{(i,i)}_{t+1}$ and therefore vanishing much faster than the other two terms as $\tilde{\textbf{V}}^{(i,i)}_{t+1}$ goes to 0, i.e.\ as the anomaly becomes stronger. 

A very similar result for rank 1 updates of determinants, the matrix determinant Lemma, can be used to show that
\begin{equation*}
\left|\hat{\bm{\Sigma}} +  \frac{\bm{\Sigma}_A^{(i,i)}}{\tilde{\textbf{V}}_{t+1}^{(i,i)}} \textbf{I}^{(i)} \right| = \left|\hat{\bm{\Sigma}}\right| \left( 1 + \frac{\bm{\Sigma}_A^{(i,i)}}{\tilde{\textbf{V}}_{t+1}^{(i,i)}} \left(\hat{\bm{\Sigma}}^{-1}\right)^{(i,i)}\right).
\end{equation*}
Furthermore, given that
\begin{align*}
-\frac{1}{2} \left(\textbf{Y}-\textbf{C}\textbf{A}\bm{\mu}\right)^T
\hat{\bm{\Sigma}}^{-1} \textbf{I}^{(j)}  \hat{\bm{\Sigma}}^{-1}
\left(\textbf{Y}-\textbf{C}\textbf{A}\bm{\mu}\right) \end{align*} 
is equal to
\begin{align*}
-\frac{1}{2} \left( \left(\hat{\bm{\Sigma}}^{-1}\right)^{(i,:)} \left(\textbf{Y}-\textbf{C}\textbf{A}\bm{\mu}\right) \right)^2,
\end{align*}
we can rewrite the posterior of $\tilde{\textbf{V}}_{t+1}^{(i,i)}$ in Equation \eqref{eq:MAIN} as \footnotesize
\begin{align*}
r_i f(\textbf{V}_{t+1}^{(i,i)})\sqrt{|\tilde{\textbf{V}}_{t+1}^{(i,i)}|}
\exp \left( 
-\frac{\tilde{\textbf{V}}_{t+1}^{(i,i)}}{2\bm{\Sigma}_A^{(i,i)}}\left( \frac{\left(\hat{\bm{\Sigma}}^{-1}\right)^{(i,:)} \left(\textbf{Y}-\textbf{C}\textbf{A}\bm{\mu}\right) }{\left(\hat{\bm{\Sigma}}^{-1}\right)^{(i,i)}}\right)^2
\right)
\frac{
	\exp \left( 
	-\frac{1}{2} \left(\textbf{Y}-\textbf{C}\textbf{A}\bm{\mu}\right)^T
	\hat{\bm{\Sigma}}^{-1} 
	\left(\textbf{Y}-\textbf{C}\textbf{A}\bm{\mu}\right)
	\right)
}{\sqrt{|\hat{\bm{\Sigma}}|}\sqrt{\left( \tilde{\textbf{V}}^{(i,i)} + \bm{\Sigma}_A^{(i,i)} \left(\hat{\bm{\Sigma}}^{-1}\right)^{(i,i)}\right)}} \\
	\exp \left( 
	\frac{1}{2} \left( 1 +\left(\frac{\tilde{\textbf{V}}_{t+1}^{(i,i)}}{\bm{\Sigma}_A^{(i,i)} \left(\hat{\bm{\Sigma}}^{-1}\right)^{(i,i)} }\right)^2 \frac{\bm{\Sigma}_A^{(i,i)} \left(\hat{\bm{\Sigma}}^{-1}\right)^{(i,i)}}{\bm{\Sigma}_A^{(i,i)} \left(\hat{\bm{\Sigma}}^{-1}\right)^{(i,i)} +  \tilde{\textbf{V}}_{t+1}^{(i,i)}} \right)
	\left( \frac{\left(\hat{\bm{\Sigma}}^{-1}\right)^{(i,:)} \left(\textbf{Y}-\textbf{C}\textbf{A}\bm{\mu}\right) }{\sqrt{\left(\hat{\bm{\Sigma}}^{-1}\right)^{(i,i)}}}\right)^2
	\right)
\end{align*} \normalsize
Using conjugacy, we can therefore sample $M$ particles for $\tilde{\textbf{V}}^{(i,i)}$ from 
\begin{equation*}
\tilde{\sigma}_i\Gamma\left(a_i + \frac{1}{2},a_i + \frac{\tilde{\sigma}_i}{2\bm{\Sigma}_A^{(i,i)}}\left( \frac{\left(\hat{\bm{\Sigma}}^{-1}\right)^{(i,:)} \left(\textbf{Y}-\textbf{C}\textbf{A}\bm{\mu}\right) }{\left(\hat{\bm{\Sigma}}^{-1}\right)^{(i,i)}}\right)^2\right)
\end{equation*}
and give each particle an importance weight proportional to
	\footnotesize
	\begin{align*}
	\frac{1}{M}r_i\frac{\Gamma(a_i+\frac{1}{2})}{\Gamma(a_i)} \sqrt{\tilde{\sigma}_i} \frac{a_i^{a_i}}{\left( 
		a_i + \frac{\tilde{\sigma}_i}{2\bm{\Sigma}_A^{(i,i)}}\left( \frac{\left(\hat{\bm{\Sigma}}^{-1}\right)^{(i,:)} \left(\textbf{Y}-\textbf{C}\textbf{A}\bm{\mu}\right) }{\left(\hat{\bm{\Sigma}}^{-1}\right)^{(i,i)}}\right)^2	
		\right)^{a_i + \frac{1}{2}}} 
	\frac{
		\exp \left( 
		-\frac{1}{2} \left(\textbf{Y}-\textbf{C}\textbf{A}\bm{\mu}\right)^T
		\hat{\bm{\Sigma}}^{-1} 
		\left(\textbf{Y}-\textbf{C}\textbf{A}\bm{\mu}\right)
		\right)
	}{\sqrt{|\hat{\bm{\Sigma}}|} \sqrt{\left( \tilde{\textbf{V}}^{(i,i)} + \bm{\Sigma}_A^{(i,i)} \left(\hat{\bm{\Sigma}}^{-1}\right)^{(i,i)}\right)}  } \\
		\exp \left( 
		\frac{1}{2} \left( 1 +\left(\frac{\tilde{\textbf{V}}_{t+1}^{(i,i)}}{\bm{\Sigma}_A^{(i,i)} \left(\hat{\bm{\Sigma}}^{-1}\right)^{(i,i)} }\right)^2 \frac{\bm{\Sigma}_A^{(i,i)} \left(\hat{\bm{\Sigma}}^{-1}\right)^{(i,i)}}{\bm{\Sigma}_A^{(i,i)} \left(\hat{\bm{\Sigma}}^{-1}\right)^{(i,i)} +  \tilde{\textbf{V}}_{t+1}^{(i,i)}} \right)
		\left( \frac{\left(\hat{\bm{\Sigma}}^{-1}\right)^{(i,:)} \left(\textbf{Y}-\textbf{C}\textbf{A}\bm{\mu}\right) }{\sqrt{\left(\hat{\bm{\Sigma}}^{-1}\right)^{(i,i)}}}\right)^2
		\right).
	\end{align*}
	\normalsize

\subsubsection{Theorem \ref{Thm:W_Anom}}\label{sec:ThmW_Anom}

\begin{Thm}\label{Thm:W_Anom}
	Let the prior for the hidden state $\textbf{X}_{t}$ be $N(\bm{\mu},\bm{\Sigma})$ and an observation $\bm{Y}_{t+1} := \bm{Y}$ be available. Then the samples for $\tilde{\textbf{W}}^{(j,j)}$ from
	\begin{equation*}
	\hat{\sigma_i}\Gamma\left(b_j + \frac{1}{2},b_j + \frac{\hat{\sigma}_j}{2\bm{\Sigma}_I^{(j,j)}}\left( \frac{\left(\textbf{C}^T\right)^{(j,:)}\hat{\bm{\Sigma}}^{-1} \left(\textbf{Y}-\textbf{C}\textbf{A}\bm{\mu}\right) }{\left(\textbf{C}^T\hat{\bm{\Sigma}}^{-1}\textbf{C}\right)^{(j,j)}}\right)^2\right)
	\end{equation*} 
	have associated weight
	\footnotesize
	\begin{align*}
	\frac{1}{M}s_j\frac{\Gamma(b_i+\frac{1}{2})}{\Gamma(b_j)} \sqrt{\hat{\sigma}_j} \frac{b_j^{b_j}}{\left( 
		b_j + \frac{\hat{\sigma}_i}{2\bm{\Sigma}_I^{(j,j)}}\left( \frac{\left(\textbf{C}^T\right)^{(j,:)}\hat{\bm{\Sigma}}^{-1} \left(\textbf{Y}-\textbf{C}\textbf{A}\bm{\mu}\right) }{\left(\textbf{C}^T\hat{\bm{\Sigma}}^{-1}\textbf{C}\right)^{(j,j)}}\right)^2	
		\right)^{b_i + \frac{1}{2}}} 
			\frac{
		\exp \left( 
		-\frac{1}{2} \left(\textbf{Y}-\textbf{C}\textbf{A}\bm{\mu}\right)^T
		\hat{\bm{\Sigma}}^{-1} 
		\left(\textbf{Y}-\textbf{C}\textbf{A}\bm{\mu}\right)
		\right)
	}{\sqrt{|\hat{\bm{\Sigma}}|} \sqrt{\left( \tilde{\textbf{W}}^{(j,j)} + \bm{\Sigma}_I^{(j,j)} \left(\textbf{C}^T\hat{\bm{\Sigma}}^{-1}\textbf{C}\right)^{(j,j)}\right)} } 
	\\
		\exp \left( 
		\frac{1}{2} \Bigg( 1 +\left(\frac{\tilde{\textbf{W}}^{(j,j)}}{\bm{\Sigma}_I^{(j,j)} \left(\textbf{C}^T\hat{\bm{\Sigma}}^{-1}\textbf{C}\right)^{(j,j)} }\right)^2  \frac{\bm{\Sigma}_I^{(j,j)} \left(\textbf{C}^T\hat{\bm{\Sigma}}^{-1}\textbf{C}\right)^{(j,j)}}{\bm{\Sigma}_I^{(j,j)} \left(\textbf{C}^T\hat{\bm{\Sigma}}^{-1}\textbf{C}\right)^{(j,j)} +  \tilde{\textbf{W}}_{t+1}^{(j,j)}} \Bigg)\left( \frac{\left(\textbf{C}^T\right)^{(j,:)}\hat{\bm{\Sigma}}^{-1} \left(\textbf{Y}-\textbf{C}\textbf{A}\bm{\mu}\right) }{\sqrt{\left(\textbf{C}^T\hat{\bm{\Sigma}}^{-1}\textbf{C}\right)^{(j,j)}}}\right)^2
		\right)
	\end{align*}
	\normalsize
\end{Thm}

The proof is almost identical to that of Theorem \ref{Thm:V_Anom} and has been omitted.

\subsubsection{Theorem \ref{Thm:No_Anom}}\label{sec:ThmNo_Anom}

\begin{Thm}\label{Thm:No_Anom}
	Let the prior for the hidden state $\textbf{X}_{t}$ be $N(\bm{\mu},\bm{\Sigma})$ and an observation $\bm{Y}_{t+1} := \bm{Y}$ be available. Then the proposal particle $(\textbf{I}_p,\textbf{I}_q)$ for $(\textbf{V}_t,\textbf{W}_t)$ has weight proportional to
	\begin{equation*}
	(1-\sum_{i=1}^{p}r_i- \sum_{j=1}^{q}s_j)
	\frac{
		\exp \left( 
		-\frac{1}{2} \left(\textbf{Y}-\textbf{C}\textbf{A}\bm{\mu}\right)^T
		\hat{\bm{\Sigma}}^{-1} 
		\left(\textbf{Y}-\textbf{C}\textbf{A}\bm{\mu}\right)
		\right)
	}{\sqrt{|\hat{\bm{\Sigma}}|}}.
	\end{equation*} 
\end{Thm}

This is immediate from the Gaussian likelihood and the Bernoulli priors for $\lambda_t^{(i)}$ and $\gamma_t^{(j)}$.

\subsubsection{Theorem \ref{Thm:Weights}}\label{sec:ThmWeights}

\begin{Thm}\label{Thm:Weights}
	Let the prior for the hidden state $\textbf{X}_{t}$ be $N(\bm{\mu},\bm{\Sigma})$ and an observation $\bm{Y}_{t+1} := \bm{Y}$ be available. When 
\begin{equation*}
\tilde{\sigma}_i = \bm{\Sigma}_A^{(i,i)} \left(\hat{\bm{\Sigma}}^{-1}\right)^{(i,i)}
\;\;
\text{and} \;\;
\hat{\sigma}_j =\bm{\Sigma}_I^{(j,j)} \left(\textbf{C}^T\hat{\bm{\Sigma}}^{-1}\textbf{C}\right)^{(j,j)},
\end{equation*}
and $a_1 = ... = a_p = b_1 = ... = b_q = c$, the weights of additive and innovative anomalies are asymptotically proportional to 
\begin{equation*}
\frac{c^{c}\frac{1}{M}r_i\frac{\Gamma(c+\frac{1}{2})}{\Gamma(c)} 
	\exp \left( 
	\frac{1}{2} \delta ^ 2 
	\right)
}{\left( 
	\frac{\delta^2}{2}
	\right)^{c}
} 
\;\;
\text{and}
\;\;
\frac{c^{c}\frac{1}{M}s_j\frac{\Gamma(c+\frac{1}{2})}{\Gamma(c)} 
	\exp \left( 
	\frac{1}{2} \delta ^ 2 
	\right)
}{\left( 
	\frac{\delta^2}{2}
	\right)^{c}
} 
\end{equation*}
when 
\small
\begin{equation*}
\textbf{Y}-\textbf{CA}\bm{\mu} = \frac{\delta \textbf{e}_i}{ \sqrt{\left( \hat{\bm{\Sigma}}^{-1}\right)^{(i,i)}} }
\;\;
\text{and} \;\;
\textbf{Y}-\textbf{C}\textbf{A}\bm{\mu} = \frac{\delta\textbf{C}^{(:,j)}}{\sqrt{\left(\textbf{C}^T\hat{\bm{\Sigma}}^{-1}\textbf{C}\right)^{(j,j)}}},
\end{equation*}
\normalsize
respectively, as $\delta \rightarrow \infty$
\end{Thm}

\textbf{Proof}: Removing the likelihood term common to all particles the importance weights can be summarised as being \footnotesize
	\begin{align*}
	\frac{1}{M}r_i\frac{\Gamma(a_i+\frac{1}{2})}{\Gamma(a_i)} \sqrt{\tilde{\sigma}_i} \frac{a_i^{a_i}}{\left( 
		a_i + \frac{\tilde{\sigma}_i}{2\bm{\Sigma}_A^{(i,i)}}\left( \frac{\left(\hat{\bm{\Sigma}}^{-1}\right)^{(i,:)} \left(\textbf{Y}-\textbf{C}\textbf{A}\bm{\mu}\right) }{\left(\hat{\bm{\Sigma}}^{-1}\right)^{(i,i)}}\right)^2	
		\right)^{a_i + \frac{1}{2}}} \frac{
	1
	}{ \sqrt{\left( \tilde{\textbf{V}}^{(i,i)} + \bm{\Sigma}_A^{(i,i)} \left(\hat{\bm{\Sigma}}^{-1}\right)^{(i,i)}\right)}  }
	\\
		\exp \left( 
		\frac{1}{2} \left( 1 +\left(\frac{\tilde{\textbf{V}}_{t+1}^{(i,i)}}{\bm{\Sigma}_A^{(i,i)} \left(\hat{\bm{\Sigma}}^{-1}\right)^{(i,i)} }\right)^2 \frac{\bm{\Sigma}_A^{(i,i)} \left(\hat{\bm{\Sigma}}^{-1}\right)^{(i,i)}}{\bm{\Sigma}_A^{(i,i)} \left(\hat{\bm{\Sigma}}^{-1}\right)^{(i,i)} +  \tilde{\textbf{V}}_{t+1}^{(i,i)}} \right)
		\left( \frac{\left(\hat{\bm{\Sigma}}^{-1}\right)^{(i,:)} \left(\textbf{Y}-\textbf{C}\textbf{A}\bm{\mu}\right) }{\sqrt{\left(\hat{\bm{\Sigma}}^{-1}\right)^{(i,i)}}}\right)^2
		\right).
	\end{align*}
	\normalsize
for the particles containing an anomaly in the $i$th additive component, and 
\footnotesize
	\begin{align*}
	\frac{1}{M}s_j\frac{\Gamma(b_i+\frac{1}{2})}{\Gamma(b_j)} \sqrt{\hat{\sigma}_j} \frac{b_j^{b_j}}{\left( 
		b_j + \frac{\hat{\sigma}_i}{2\bm{\Sigma}_I^{(j,j)}}\left( \frac{\left(\textbf{C}^T\right)^{(j,:)}\hat{\bm{\Sigma}}^{-1} \left(\textbf{Y}-\textbf{C}\textbf{A}\bm{\mu}\right) }{\left(\textbf{C}^T\hat{\bm{\Sigma}}^{-1}\textbf{C}\right)^{(j,j)}}\right)^2	
		\right)^{b_i + \frac{1}{2}}} 
		\frac{
	1
	}{ \sqrt{\left( \tilde{\textbf{W}}^{(j,j)} + \bm{\Sigma}_I^{(j,j)} \left(\textbf{C}^T\hat{\bm{\Sigma}}^{-1}\textbf{C}\right)^{(j,j)}\right)} } \\
	\exp \Bigg( 
		\frac{1}{2} \left( 1 +\left(\frac{\tilde{\textbf{W}}^{(j,j)}}{\bm{\Sigma}_I^{(j,j)} \left(\textbf{C}^T\hat{\bm{\Sigma}}^{-1}\textbf{C}\right)^{(j,j)} }\right)^2  \frac{\bm{\Sigma}_I^{(j,j)} \left(\textbf{C}^T\hat{\bm{\Sigma}}^{-1}\textbf{C}\right)^{(j,j)}}{\bm{\Sigma}_I^{(j,j)} \left(\textbf{C}^T\hat{\bm{\Sigma}}^{-1}\textbf{C}\right)^{(j,j)} +  \tilde{\textbf{W}}_{t+1}^{(j,j)}} \Bigg)\left( \frac{\left(\textbf{C}^T\right)^{(j,:)}\hat{\bm{\Sigma}}^{-1} \left(\textbf{Y}-\textbf{C}\textbf{A}\bm{\mu}\right) }{\sqrt{\left(\textbf{C}^T\hat{\bm{\Sigma}}^{-1}\textbf{C}\right)^{(j,j)}}}\right)^2
		\right)
	\end{align*}
\normalsize
for the particles containing an anomaly in the $j$th innovative component. 

As mentioned in Section II that the mean of the proposal of the $i$th additive component behaves asymptotically as
\begin{equation*}
    (2a_i+1)\bm{\Sigma}_A^{(i,i)} \left( \frac{\left(\hat{\bm{\Sigma}}^{-1}\right)^{(i,i)}}{\left(\hat{\bm{\Sigma}}^{-1}\right)^{(i,:)} \left(\textbf{Y}-\textbf{C}\textbf{A}\bm{\mu}\right) }\right)^2.
\end{equation*}
Furthermore, the standard deviation is on the same scale. We therefore have that 
\begin{equation*}
    \tilde{\textbf{V}}^{(i,i)}_{t+1} \sim \frac{1}{\delta^2}
\end{equation*}
as $\delta \rightarrow \infty$. The weight of an anomaly in the $i$th additive component therefore asymptotically behaves as
\begin{align*}
\frac{a_i^{a_i}\frac{1}{M}r_i\frac{\Gamma(a_i+\frac{1}{2})}{\Gamma(a_i)} 
	\exp \left( 
	\frac{1}{2} \delta ^ 2 
	\right)
}{\left( 
	\frac{\tilde{\sigma}_i}{2\bm{\Sigma}_A^{(i,i)}\left(\hat{\bm{\Sigma}}^{-1}\right)^{(i,i)}  }\delta^2
	\right)^{a_i }
} 
\end{align*}
when $\textbf{Y}-\textbf{CA}\bm{\mu} = \frac{1}{ \sqrt{\left( \hat{\bm{\Sigma}}^{-1}\right)^{(i,i)}} }\delta \textbf{e}_i  $ as $\delta \rightarrow \infty$. A very similar reasoning can be used to show that the weight of an anomaly in the $j$th innovative component converges to
\begin{equation*}
\frac{b_j^{b_j}\frac{1}{M}s_j\frac{\Gamma(b_j+\frac{1}{2})}{\Gamma(b_j)} 
	\exp \left( 
	\frac{1}{2} \delta ^ 2 
	\right)
}{\left( 
	\frac{\hat{\sigma}_j}{2\bm{\Sigma}_I^{(j,j)} 
		\left(\textbf{C}^T\hat{\bm{\Sigma}}^{-1}\textbf{C}\right)^{(j,j)}	
	}\delta^2
	\right)^{b_j}
} 
\end{equation*}
when $\textbf{Y}-\textbf{C}\textbf{A}\bm{\mu} = \frac{\textbf{C}^{(:,ij}}{\sqrt{\left(\textbf{C}^T\hat{\bm{\Sigma}}^{-1}\textbf{C}\right)^{(j,j)}}}\delta $ as $\delta \rightarrow \infty$.  

The result then follows when all the $b_j$s and the $a_i$s are equal to the same constant $c$ and 
\begin{equation*}
\tilde{\sigma}_i = \bm{\Sigma}_A^{(i,i)} \left(\hat{\bm{\Sigma}}^{-1}\right)^{(i,i)}
\;\;\;
\text{and} \;\;\;
\hat{\sigma}_j =\bm{\Sigma}_I^{(j,j)} \left(\textbf{C}^T\hat{\bm{\Sigma}}^{-1}\textbf{C}\right)^{(j,j)}.
\end{equation*}

\subsubsection{Theorem \ref{Thm:BS}}

\begin{Thm}\label{Thm:BS}
	Let the prior for the hidden state $\textbf{X}_{t-k}$ be $N(\bm{\mu},\bm{\Sigma})$. Then the samples for $\tilde{\textbf{W}}^{(j,j)}_{t-k+1}$ from
	\footnotesize
	\begin{equation*}
\hat{\sigma_j}\Gamma\left(b_j + \frac{1}{2},b_j + \frac{\hat{\sigma_j}}{2\bm{\Sigma}_I^{(j,j)}}\left( \frac{\left(\left(\tilde{\textbf{C}}^{(k)}\right)^T\right)^{(j,:)}\left(\hat{\bm{\Sigma}}^{(k)}\right)^{-1} \tilde{\textbf{z}}_{t+1-k}^{(k)} }{\left(\left(\tilde{\textbf{C}}^{(k)}\right)^T\left(\hat{\bm{\Sigma}}^{(k)}\right)^{-1}\tilde{\textbf{C}}^{(k)}\right)^{(j,j)}}\right)^2\right),
	\end{equation*} 
	\normalsize
 where $ \tilde{\textbf{z}}_{t+1-k}^{(k)} =  \tilde{\textbf{Y}}_{t+1-k}^{(k)}-\tilde{\textbf{C}}^{(k)} \textbf{A}\bm{\mu}$ have associated weight
\tiny
\begin{align*}
 \frac{\frac{1}{M}s_i\left(1 - \sum_{i'=1}^{p}r_{i'} -  \sum_{j'=1}^{q}s_{j'}\right)^{k}\frac{\Gamma(b_j+\frac{1}{2})}{\Gamma(b_j)} \sqrt{\hat{\sigma}_j} b_j^{b_j} }{\left( 
	b_i + \frac{\hat{\sigma}_j}{2\bm{\Sigma}_I^{(j,j)}}\left( \frac{\left(\left(\tilde{\textbf{C}}^{(k)}\right)^T\right)^{(j,:)}\left(\hat{\bm{\Sigma}}^{(k)}\right)^{-1} \left(\tilde{\textbf{z}}_{t+1-k}^{(k)}\right) }{\left(\left(\tilde{\textbf{C}}^{(k)}\right)^T\left(\hat{\bm{\Sigma}}^{(k)}\right)^{-1}\tilde{\textbf{C}}^{(k)}\right)^{(j,j)}}\right)^2	
	\right)^{b_j + \frac{1}{2}}} 
\frac{\exp \left( 
 	-\frac{1}{2} \left(\tilde{\textbf{z}}_{t+1-k}^{(k)}\right)^T
 	\left(\hat{\bm{\Sigma}}^{(k)}\right)^{-1}
 	\left(\tilde{\textbf{z}}_{t+1-k}^{(k)}\right)
 	\right)}{\sqrt{\left|\hat{\bm{\Sigma}}^{(k)}\right|}\sqrt{\left( \textbf{W}^{(j,j)} + \bm{\Sigma}_I^{(j,j)} \left(\left(\tilde{\textbf{C}}^{(k)}\right)^T\left(\hat{\bm{\Sigma}}^{(k)}\right)^{-1}\left(\tilde{\textbf{C}}^{(k)}\right)\right)^{(j,j)}\right)}}
\\
	\exp \Bigg( 
	\frac{1}{2} \left( 1 +\left(\frac{\textbf{W}_{t+1}^{(j,j)}}{\bm{\Sigma}_I^{(j,j)} \left(\left(\tilde{\textbf{C}}^{(k)}\right)^T\left(\hat{\bm{\Sigma}}^{(k)}\right)^{-1}\left(\tilde{\textbf{C}}^{(k)}\right)\right)^{(j,j)} }\right)^2 \frac{\bm{\Sigma}_I^{(j,j)} \left(\left(\tilde{\textbf{C}}^{(k)}\right)^T\left(\hat{\bm{\Sigma}}^{(k)}\right)^{-1}\left(\tilde{\textbf{C}}^{(k)}\right)\right)^{(j,j)}}{\bm{\Sigma}_I^{(j,j)} \left(\left(\tilde{\textbf{C}}^{(k)}\right)^T\left(\hat{\bm{\Sigma}}^{(k)}\right)^{-1}\left(\tilde{\textbf{C}}^{(k)}\right)\right)^{(j,j)} +  \textbf{W}_{t+1}^{(j,j)}} \right)\\ \left( \frac{\left(\left(\tilde{\textbf{C}}^{(k)}\right)^T\right)^{(j,:)}\left(\hat{\bm{\Sigma}}^{(k)}\right)^{-1} \left(\tilde{\textbf{Y}}_{t+1-k}^{(k)}-\left(\tilde{\textbf{C}}^{(k)}\right)\textbf{A}\bm{\mu}_{t-k}\right) }{\sqrt{\left(\left(\tilde{\textbf{C}}^{(k)}\right)^T\left(\hat{\bm{\Sigma}}^{(k)}\right)^{-1}\left(\tilde{\textbf{C}}^{(k)}\right)\right)^{(j,j)}}}\right)^2
	\Bigg)
\end{align*}
\normalsize 
\end{Thm}

\textbf{Proof}: Identical (up to variable names) to that of Theorem \ref{Thm:W_Anom}.

\subsection{Additional Simulations}

\begin{figure} 
	\begin{subfigure}[b]{0.24\linewidth}
		\centering
		\includegraphics[width=0.95\linewidth]{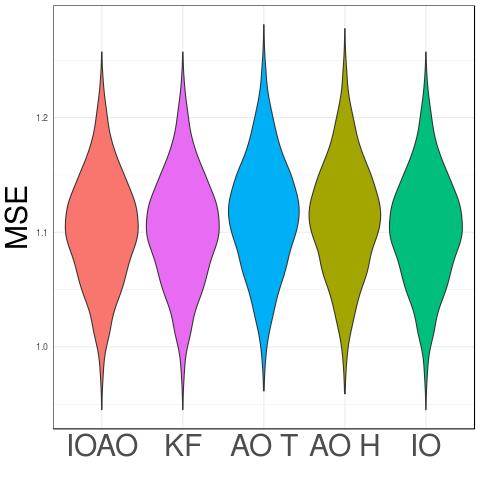} 
		\caption{Case 1}
		\label{fig:meanchange_graph} 
	\end{subfigure} 
	\begin{subfigure}[b]{0.24\linewidth}
		\centering
		\includegraphics[width=0.95\linewidth]{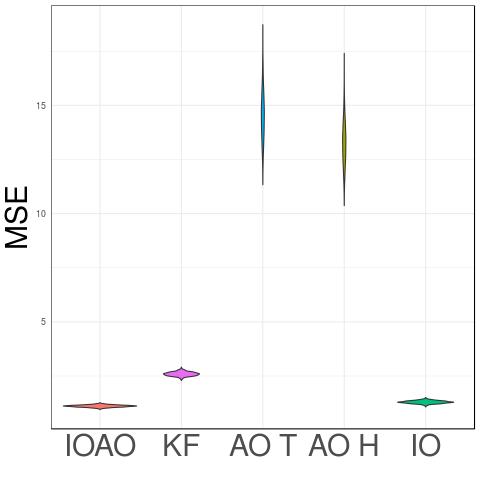} 
		\caption{Case 1, IOs} 
		\label{fig:meanANOMchange_graph} 
	\end{subfigure} 
		\begin{subfigure}[b]{0.24\linewidth}
		\centering
		\includegraphics[width=0.95\linewidth]{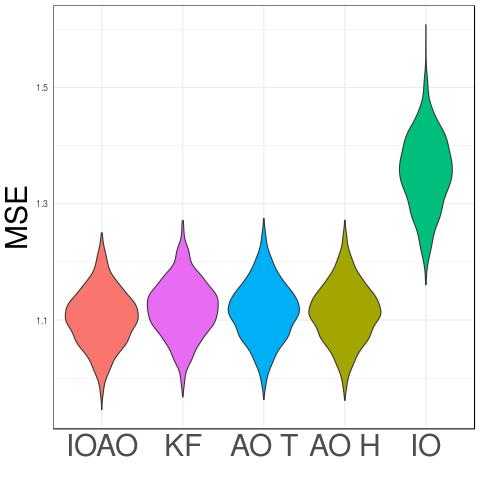} 
		\caption{Case 1, AOs} 
		\label{fig:meanANOMchange_graph} 
	\end{subfigure} 
	\begin{subfigure}[b]{0.24\linewidth}
		\centering
		\includegraphics[width=0.95\linewidth]{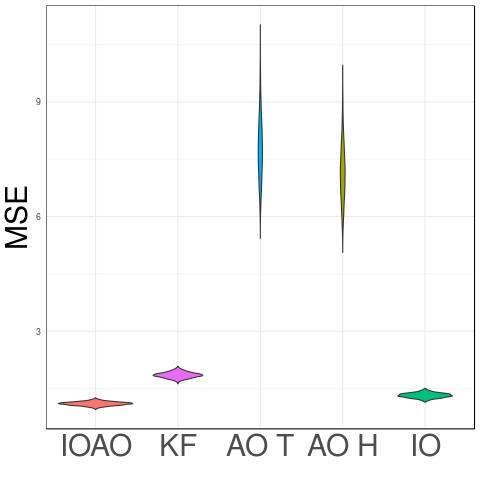}
		\caption{Case 1, Both}
		\label{fig:meanchange_graph} 
	\end{subfigure}
		\begin{subfigure}[b]{0.24\linewidth}
		\centering
		\includegraphics[width=0.95\linewidth]{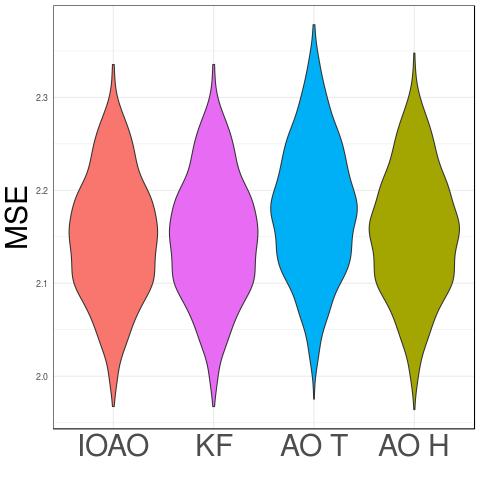} 
		\caption{Case 2}
		\label{fig:meanchange_graph} 
	\end{subfigure} 
	\begin{subfigure}[b]{0.24\linewidth}
		\centering
		\includegraphics[width=0.95\linewidth]{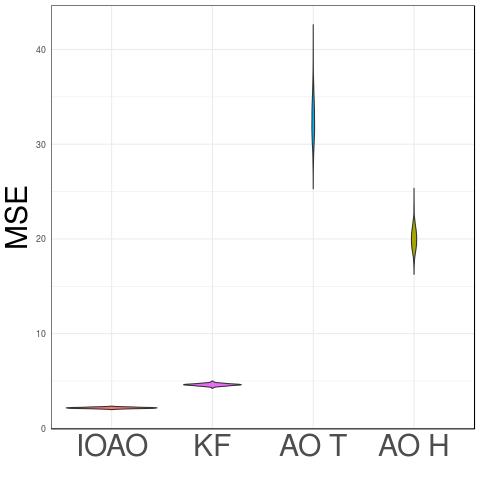} 
		\caption{Case 2, IOs} 
		\label{fig:meanANOMchange_graph} 
	\end{subfigure} 
		\begin{subfigure}[b]{0.24\linewidth}
		\centering
		\includegraphics[width=0.95\linewidth]{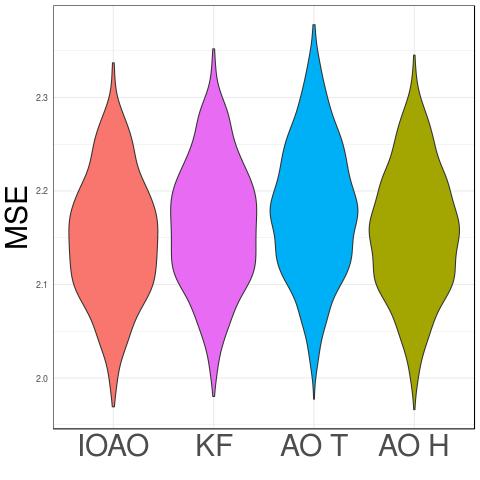} 
		\caption{Case 2, AOs} 
		\label{fig:meanANOMchange_graph} 
	\end{subfigure} 
	\begin{subfigure}[b]{0.24\linewidth}
		\centering
		\includegraphics[width=0.95\linewidth]{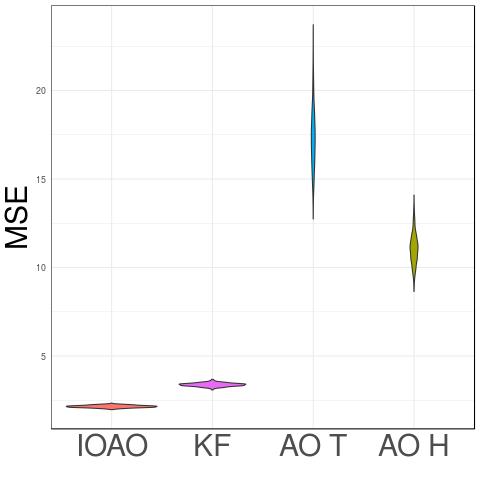}
		\caption{Case 2, Both}
		\label{fig:meanchange_graph} 
	\end{subfigure}
			\begin{subfigure}[b]{0.24\linewidth}
		\centering
		\includegraphics[width=0.95\linewidth]{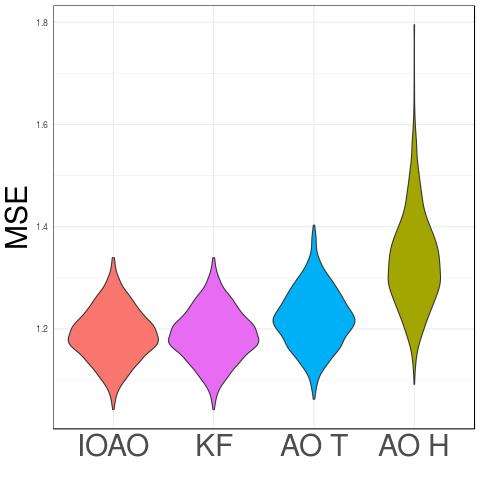} 
		\caption{Case 3}
		\label{fig:meanchange_graph} 
	\end{subfigure} 
	\begin{subfigure}[b]{0.24\linewidth}
		\centering
		\includegraphics[width=0.95\linewidth]{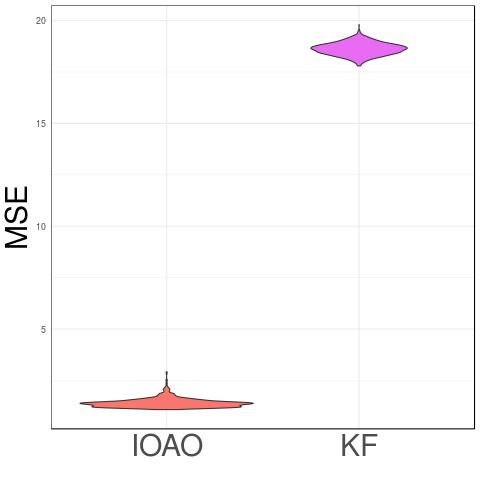} 
		\caption{Case 3, IOs} 
		\label{fig:meanANOMchange_graph} 
	\end{subfigure} 
		\begin{subfigure}[b]{0.24\linewidth}
		\centering
		\includegraphics[width=0.95\linewidth]{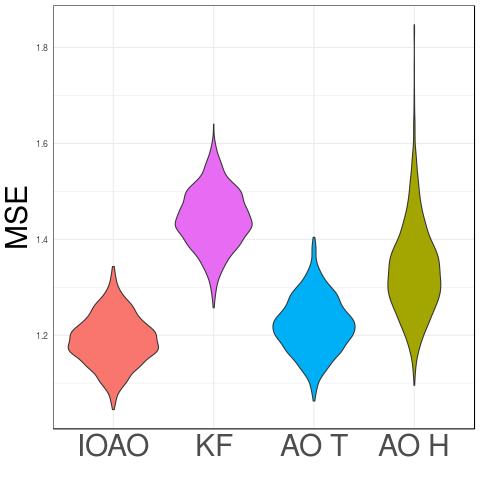} 
		\caption{Case 3, AOs} 
		\label{fig:meanANOMchange_graph} 
	\end{subfigure} 
	\begin{subfigure}[b]{0.245\linewidth}
		\centering
		\includegraphics[width=0.95\linewidth]{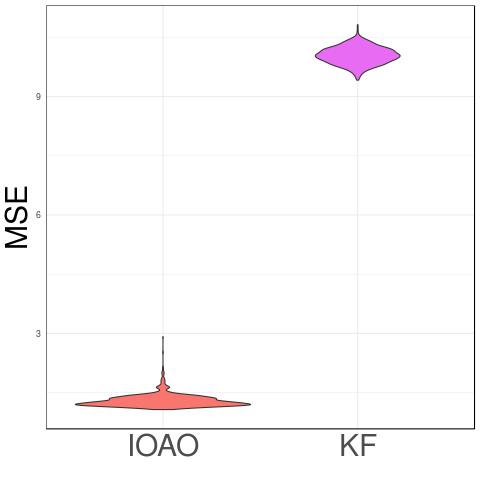}
		\caption{Case 3, Both}
		\label{fig:meanchange_graph} 
	\end{subfigure}
			\begin{subfigure}[b]{0.24\linewidth}
		\centering
		\includegraphics[width=0.95\linewidth]{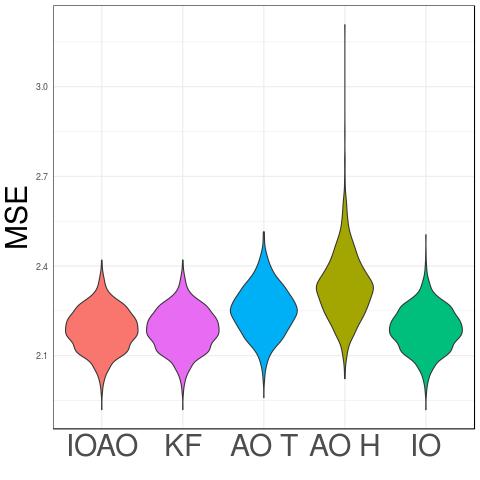} 
		\caption{Case 4}
		\label{fig:meanchange_graph} 
	\end{subfigure} 
	\begin{subfigure}[b]{0.24\linewidth}
		\centering
		\includegraphics[width=0.95\linewidth]{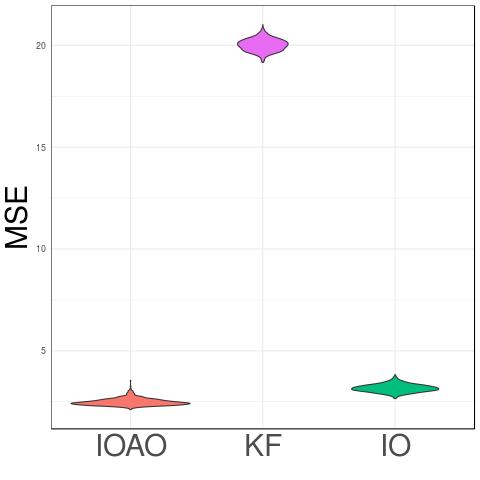} 
		\caption{Case 4, IOs} 
		\label{fig:meanANOMchange_graph} 
	\end{subfigure} 
		\begin{subfigure}[b]{0.24\linewidth}
		\centering
		\includegraphics[width=0.95\linewidth]{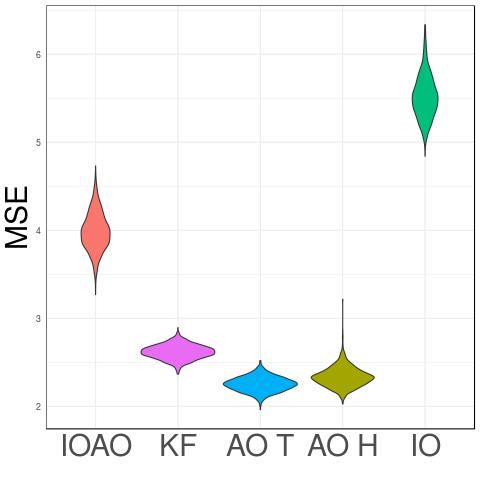} 
		\caption{Case 4, AOs} 
		\label{fig:meanANOMchange_graph} 
	\end{subfigure} 
	\begin{subfigure}[b]{0.24\linewidth}
		\centering
		\includegraphics[width=0.95\linewidth]{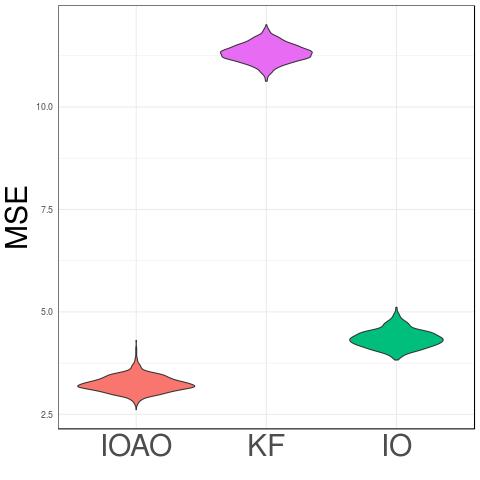}
		\caption{Case 4, Both}
		\label{fig:meanchange_graph} 
	\end{subfigure}
	\caption{Violin plots for the average predictive mean squared error of the five filters over the four different scenarios under a range of models. Lower values correspond to better performance. Methods are omitted if they can not be applied to the setting or if their performance is too poor.}
	\label{fig:MSE} 
\end{figure}

Violin plots for the predictive mean squared error are displayed in Figure \ref{fig:MSE}

\subsection{Complete pseudocode}

\setcounter{algorithm}{2}

\begin{algorithm}
	\caption{KF\_Upd($\textbf{Y},\bm{\mu},\bm{\Sigma},\textbf{C},\textbf{A},\bm{\Sigma}_A,\bm{\Sigma}_I$)}
	\begin{footnotesize}
		\begin{algorithmic}[1]
		\State ${\bm{\mu}}_p \gets \textbf{A} \bm{\mu}$
		\State ${\bm{\Sigma}}_p \gets \textbf{A} \bm{\Sigma} \textbf{A}^T + \bm{\Sigma}_I$
		\State $\textbf{z} = \textbf{Y} - {\bm{\mu}}_p$
		\State $\hat{\bm{\Sigma}} \gets \textbf{C} \bm{\Sigma}_p \textbf{C}^T + \bm{\Sigma}_A$
		\State $\textbf{K} \gets \bm{\Sigma}_p \textbf{C}^T \hat{\bm{\Sigma}}^{-1}$
		\State $\bm{\mu}_{new} \gets \bm{\mu}_p + \textbf{K} \textbf{z}$
		\State $\bm{\Sigma}_{new} \gets \left(\textbf{I} - \textbf{K} \textbf{C}\right){\bm{\Sigma}}_p$
	\end{algorithmic}
	\begin{tabular}[h]{ll}
		{\bf Output:} & $(\bm{\mu}_{new},\bm{\Sigma}_{new})$
	\end{tabular}
	\end{footnotesize}
\end{algorithm}

\begin{algorithm}
	\caption{Sample\_typical($\bm{\mu},\bm{\Sigma},\textbf{Y},\textbf{A},\textbf{C},\bm{\Sigma}_A,\bm{\Sigma}_I$)}
	\begin{footnotesize}
		\begin{algorithmic}[1]
			\State $\textbf{V} \gets \textbf{I}_p$
			\State $\textbf{W} \gets \textbf{I}_q$
			\State $\hat{\bm{\Sigma}} \gets \textbf{C}\left( \textbf{A} \bm{\Sigma} \textbf{A}^T +\bm{\Sigma}_I\right) \textbf{C}^T + \bm{\Sigma}_A $
			\State $\textbf{z}\gets \textbf{Y} - \textbf{C} \textbf{A} \bm{\mu}$
			\State $prob \gets \left(1 - \sum_{i=1}^pr_i - \sum_{j=1}^qs_j \right)\exp \left( - \frac{1}{2} \textbf{z}^T \hat{\bm{\Sigma}} ^{-1}\textbf{z} \right) / \sqrt{ \left|  \hat{\bm{\Sigma}} \right|}$
		\end{algorithmic}
		\begin{tabular}[h]{ll}
			{\bf Output:} & $(\textbf{V},\textbf{W},prob)$
		\end{tabular}
	\end{footnotesize}
\end{algorithm}

\begin{algorithm}
	\caption{Sample\_add\_comp($i,\textbf{z},\hat{\bm{\Sigma}},\bm{\Sigma}_A,M$)}
	\begin{footnotesize}
		\begin{algorithmic}[1]
			\State $\textbf{V} \gets \textbf{I}_p $
			\State $\textbf{V} \gets \textbf{I}_q $
			\State $\textbf{V}^{(i,i)} \gets \tilde{\sigma}_i\Gamma\left(a_i + \frac{1}{2},a_i + \frac{\tilde{\sigma}_i}{2\bm{\Sigma}_A^{(i,i)}}\left( \frac{\left(\hat{\bm{\Sigma}}^{-1}\right)^{(i,:)} \textbf{z} }{\left(\hat{\bm{\Sigma}}^{-1}\right)^{(i,i)}}\right)^2\right)$
			\State 
			\footnotesize
			\begin{align*}
			    prob \gets \frac{1}{M}r_i\frac{\Gamma(a_i+\frac{1}{2})}{\Gamma(a_i)} \frac{a_i^{a_i}}{\left( 
		a_i + \frac{\tilde{\sigma}_i}{2\bm{\Sigma}_A^{(i,i)}}\left( \frac{\left(\hat{\bm{\Sigma}}^{-1}\right)^{(i,:)}\textbf{z} }{\left(\hat{\bm{\Sigma}}^{-1}\right)^{(i,i)}}\right)^2	
		\right)^{a_i + \frac{1}{2}}} 
	\frac{
		\sqrt{\tilde{\sigma}_i} \exp \left( 
		-\frac{1}{2} \textbf{z}^T
		\hat{\bm{\Sigma}}^{-1} 
		\textbf{z}
		\right)
	}{\sqrt{|\hat{\bm{\Sigma}}|} \sqrt{\left( \tilde{\textbf{V}}^{(i,i)} + \bm{\Sigma}_A^{(i,i)} \left(\hat{\bm{\Sigma}}^{-1}\right)^{(i,i)}\right)}  } \\
		\exp \left( 
		\frac{1}{2} \left( 1 +\left(\frac{\tilde{\textbf{V}}_{t+1}^{(i,i)}}{\bm{\Sigma}_A^{(i,i)} \left(\hat{\bm{\Sigma}}^{-1}\right)^{(i,i)} }\right)^2 \frac{\bm{\Sigma}_A^{(i,i)} \left(\hat{\bm{\Sigma}}^{-1}\right)^{(i,i)}}{\bm{\Sigma}_A^{(i,i)} \left(\hat{\bm{\Sigma}}^{-1}\right)^{(i,i)} +  \tilde{\textbf{V}}_{t+1}^{(i,i)}} \right)
		\left( \frac{\left(\hat{\bm{\Sigma}}^{-1}\right)^{(i,:)} \textbf{z} }{\sqrt{\left(\hat{\bm{\Sigma}}^{-1}\right)^{(i,i)}}}\right)^2
		\right).\end{align*}
		\end{algorithmic}
		\begin{tabular}[h]{ll}
			{\bf Output:} & $(\textbf{V},\textbf{W},prob)$
		\end{tabular}
		\normalsize
	\end{footnotesize}
\end{algorithm}

\begin{algorithm}
	\caption{Sample\_add($\bm{\mu},\bm{\Sigma},\textbf{Y},\textbf{A},\textbf{C},\bm{\Sigma}_A,\bm{\Sigma}_I,M$)}
	\begin{footnotesize}
		\begin{algorithmic}[1]
			\State $\hat{\bm{\Sigma}} \gets \textbf{C}\left( \textbf{A} \bm{\Sigma} \textbf{A}^T +\bm{\Sigma}_I\right) \textbf{C}^T + \bm{\Sigma}_A $
			\State $\textbf{z}\gets \textbf{Y} - \textbf{C} \textbf{A} \bm{\mu}$
			\State $Add\_Pt  \gets \{ \}$ \Comment{Additive Anom. Particles}
			\For {$ i \in \{1,...,p\}$}
			\State $Add\_Pt \gets Add\_Pt \cup \{\text{Sample\_add\_comp}(i,\textbf{z},\hat{\bm{\Sigma}},\bm{\Sigma}_A,M)\}$
			\EndFor
		\end{algorithmic}
		\begin{tabular}[h]{ll}
			{\bf Output:} & $Add\_Pt$
		\end{tabular}
	\end{footnotesize}
\end{algorithm}

\begin{algorithm}
	\caption{Sample\_inn\_comp($j,\textbf{z},\hat{\bm{\Sigma}},\bm{\Sigma}_I,M$)}
	\begin{footnotesize}
		\begin{algorithmic}[1]
			\State $\textbf{V} \gets \textbf{I}_p $
			\State $\textbf{V} \gets \textbf{I}_q $
			\State $\textbf{W}^{(i,i)} \gets \hat{\sigma_i}\Gamma\left(b_i + \frac{1}{2},b_i + \frac{\hat{\sigma_i}}{2\bm{\Sigma}_I^{(i,i)}}\left( \frac{\left(\textbf{C}^T\right)^{(i,:)}\hat{\bm{\Sigma}}^{-1} \textbf{z} }{\left(\textbf{C}^T\hat{\bm{\Sigma}}^{-1}\textbf{C}\right)^{(i,i)}}\right)^2\right)$ 
				\footnotesize
					\State
	\begin{align*}
 prob \gets \frac{1}{M}s_j\frac{\Gamma(b_i+\frac{1}{2})}{\Gamma(b_j)} \frac{b_j^{b_j}}{\left( 
		b_j + \frac{\hat{\sigma}_i}{2\bm{\Sigma}_I^{(j,j)}}\left( \frac{\left(\textbf{C}^T\right)^{(j,:)}\hat{\bm{\Sigma}}^{-1} \textbf{z} }{\left(\textbf{C}^T\hat{\bm{\Sigma}}^{-1}\textbf{C}\right)^{(j,j)}}\right)^2	
		\right)^{b_i + \frac{1}{2}}} 
			\frac{
		\sqrt{\hat{\sigma}_j}  \exp \left( 
		-\frac{1}{2} \textbf{z}^T
		\hat{\bm{\Sigma}}^{-1} 
		\textbf{z}
		\right)
	}{\sqrt{|\hat{\bm{\Sigma}}|} \sqrt{\left( \tilde{\textbf{W}}^{(j,j)} + \bm{\Sigma}_I^{(j,j)} \left(\textbf{C}^T\hat{\bm{\Sigma}}^{-1}\textbf{C}\right)^{(j,j)}\right)} } 
	\\
		\exp \left( 
		\frac{1}{2} \Bigg( 1 +\left(\frac{\tilde{\textbf{W}}^{(j,j)}}{\bm{\Sigma}_I^{(j,j)} \left(\textbf{C}^T\hat{\bm{\Sigma}}^{-1}\textbf{C}\right)^{(j,j)} }\right)^2  \frac{\bm{\Sigma}_I^{(j,j)} \left(\textbf{C}^T\hat{\bm{\Sigma}}^{-1}\textbf{C}\right)^{(j,j)}}{\bm{\Sigma}_I^{(j,j)} \left(\textbf{C}^T\hat{\bm{\Sigma}}^{-1}\textbf{C}\right)^{(j,j)} +  \tilde{\textbf{W}}_{t+1}^{(j,j)}} \Bigg)\left( \frac{\left(\textbf{C}^T\right)^{(j,:)}\hat{\bm{\Sigma}}^{-1} \textbf{z} }{\sqrt{\left(\textbf{C}^T\hat{\bm{\Sigma}}^{-1}\textbf{C}\right)^{(j,j)}}}\right)^2
		\right)
	\end{align*}
		\end{algorithmic}
		\begin{tabular}[h]{ll}
			{\bf Output:} & $(\textbf{V},\textbf{W},prob)$
		\end{tabular}
	\end{footnotesize}
	\normalsize
\end{algorithm}

\begin{algorithm}
	\caption{Sample\_inn($\bm{\mu},\bm{\Sigma},\textbf{Y},\textbf{A},\textbf{C},\bm{\Sigma}_A,\bm{\Sigma}_I,M$)}
	\begin{footnotesize}
		\begin{algorithmic}[1]
			\State $\hat{\bm{\Sigma}} \gets \textbf{C}\left( \textbf{A} \bm{\Sigma} \textbf{A}^T +\bm{\Sigma}_I\right) \textbf{C}^T + \bm{\Sigma}_A $
			\State $\textbf{z}\gets \textbf{Y} - \textbf{C} \textbf{A} \bm{\mu}$
			\State $Inn\_Pt  \gets \{ \}$ \Comment{Innovative Anom. Particles}
			\For {$ i \in \{1,...,q\}$}
			\State $Inn\_Pt \gets Inn\_Pt \cup \{\text{Sample\_inn\_comp}(i,\textbf{z},\hat{\bm{\Sigma}},\bm{\Sigma}_I,M)\}$
			\EndFor
		\end{algorithmic}
		\begin{tabular}[h]{ll}
			{\bf Output:} & $Inn\_Pt$
		\end{tabular}
	\end{footnotesize}
\end{algorithm}

\begin{algorithm}
	\caption{Sample\_Particles($M,\bm{\mu},\bm{\Sigma},\textbf{Y},\textbf{A},\textbf{C},\bm{\Sigma}_A,\bm{\Sigma}_I$)}
	\begin{footnotesize}
		\begin{algorithmic}[1]
			\State $Desc \gets \{\}$ \Comment{To store Descendants}
			\State $Desc \gets Desc \cup \text{Sample\_typical}(\bm{\mu},\bm{\Sigma},\textbf{Y},\textbf{A},\textbf{C},\bm{\Sigma}_A,\bm{\Sigma}_I)$
			\For {$i \in {1,...,M}$}
			\State $Desc \gets Desc \cup \text{Sample\_add}(\bm{\mu},\bm{\Sigma},\textbf{Y},\textbf{A},\textbf{C},\bm{\Sigma}_A,\bm{\Sigma}_I,M)$
			\EndFor
			\For {$i \in {1,...,M}$}
			\State $Desc \gets Desc \cup \text{Sample\_inn}(\bm{\mu},\bm{\Sigma},\textbf{Y},\textbf{A},\textbf{C},\bm{\Sigma}_A,\bm{\Sigma}_I,M)$
			\EndFor
		\end{algorithmic}
		\begin{tabular}[h]{ll}
			{\bf Output:} & $Desc$
		\end{tabular}
	\end{footnotesize}
\end{algorithm}

\begin{algorithm}
	\caption{BS\_inn $(\bm{\mu},\bm{\Sigma},\tilde{\textbf{Y}},\textbf{A},\textbf{C},\bm{\Sigma}_A,\bm{\Sigma}_I,M,horizon)$}
	\begin{footnotesize}
		
		\begin{algorithmic}[1]
			\State $\tilde{\textbf{C}} \gets \textbf{C} \left[\left(\textbf{A}^0\right)^T,...,\left(\textbf{A}^{horizon}\right)^T\right]^T$
			\State $\tilde{\textbf{z}} \gets \tilde{\textbf{Y}} - \tilde{\textbf{C}} \textbf{A} \bm{\mu}$
			\State $\tilde{\bm{\Sigma}} \gets\tilde{\textbf{C}} \left( \textbf{A} \bm{\Sigma} \textbf{A}^T + \textbf{I}_{horizon}\otimes\bm{\Sigma}_I \right)\tilde{\textbf{C}}^T +\textbf{I}_{horizon}\otimes\bm{\Sigma}_A $
			\State $Cd\gets \{\}$ \Comment{To store Candidates.}
			\For{$i \in \{1,..,q\} $}
			\If {$horizon \in \mathcal{B}_i$}
			\For{$j \in \{1,...,M\}$}
			\State $Cd \gets Cd \cup \{\text{Sample\_inn\_comp}(i,\tilde{\textbf{z}},\tilde{\bm{\Sigma}},\textbf{A},\tilde{\textbf{C}},\bm{\Sigma}_I,M\cdot |\mathcal{B}_i|)\}$ 
			\EndFor
			\EndIf
			\EndFor
		\end{algorithmic}
		\begin{tabular}[h]{ll}
			{\bf Output:} & $Cand$
		\end{tabular}
	\end{footnotesize}
\end{algorithm}

\setcounter{algorithm}{0}

\begin{algorithm}
	\caption{Basic Particle Filter (No Back-sampling)}
	\label{alg:Basic}
	\begin{footnotesize}
		\begin{tabular}[h]{ll}
			{\bf Input:} & An initial state estimate $(\bm{\mu}_0,\bm{\Sigma}_0)$ \\ & A number of descendants, $M'=M(p+q)+1$ \\ & A number of particles to be maintained, $N$.
			 \\ & A stream of observations $\textbf{Y}_1,\textbf{Y}_2,...$ \\ {\bf Initialise:} & Set $Particles(0) = \{(\bm{\mu}_0,\bm{\Sigma}_0)\}$ 
		\end{tabular}

		\begin{algorithmic}[1]
			\For{$t \in \mathbb{N}^+ $}
			\State $Candidates \gets \{\}$ 
			\For{$(\bm{\mu},\bm{\Sigma}) \in Particles(t-1)$}
			\State $(\textbf{V},\textbf{W},prob) \gets \text{Sample\_Particles}(M,\bm{\mu},\bm{\Sigma},\textbf{Y}_t,\textbf{A},\textbf{C},\bm{\Sigma}_A,\bm{\Sigma}_I)$
			\State $Candidates \gets Candidates \cup \{(\bm{\mu},\bm{\Sigma},\textbf{V},\textbf{W},prob)\}$ 
			\EndFor
			\State $Descendants \gets \text{Subsample}(N,Candidates)$
			\State $Particles(t) \gets \{\}$
			\For{$(\bm{\mu},\bm{\Sigma},\textbf{V},\textbf{W},prob) \in Descendants $}
			\State $(\bm{\mu}_{new},\bm{\Sigma}_{new}) \gets  \text{KF\_Upd}(\textbf{Y}_t,\bm{\mu},\bm{\Sigma},\textbf{C},\textbf{A},\textbf{V}^{1/2}\bm{\Sigma}_A,\textbf{W}^{1/2}\bm{\Sigma}_I)$
			\State $Particles(t) \gets Particles(t) \cup \{(\bm{\mu}_{new},\bm{\Sigma}_{new}) \}$
			\EndFor
			\EndFor
		\end{algorithmic}
	\end{footnotesize}
\end{algorithm}
\begin{algorithm}
	\caption{Particle Filter (With Back Sampling) -- CE-BASS}
	\label{alg:Back-sample}
	\begin{footnotesize}
		\begin{tabular}[h]{ll}
			{\bf Input:} & An initial state estimate $(\bm{\mu}_0,\bm{\Sigma}_0)$. \\ & A number of descendants, $M'=M(p+q)+1$. \\ & A number of particles to be maintained, $N$.
			\\ & A stream of observations $\textbf{Y}_1,\textbf{Y}_2,...$ \\ {\bf Initialise:} & Set $Particles(0) = \{(\bm{\mu}_0,\bm{\Sigma}_0,1)\}$ \\
			& Set $max\_horizon = \max \left(\cup_{i=1}^q \mathcal{B}_i\right)$
		\end{tabular}
		
		\begin{algorithmic}[1]
			\For{$t \in \mathbb{N}^+ $}
			\State $Cand \gets \{\}$ \Comment{To Store Candidates}
			\For{$(\bm{\mu},\bm{\Sigma},prob_{prev}) \in Particles(t-1)$}
			\State $(\textbf{V},\textbf{W},prob) \gets  \text{Sample\_typical}(\bm{\mu},\bm{\Sigma},\textbf{Y}_t,\textbf{A},\textbf{C},\bm{\Sigma}_A,\bm{\Sigma}_I)$
			\State $Cand \gets Cand \cup \{(\bm{\mu},\bm{\Sigma},\textbf{V},\textbf{W},prob\cdot prob_{prev},1)\}$
			\State $Add\_Des \gets \text{Sample\_add}(\bm{\mu},\bm{\Sigma},\textbf{Y}_t,\textbf{A},\textbf{C},\bm{\Sigma}_A,\bm{\Sigma}_I,M)$
			\For {$(\textbf{V},\textbf{W},prob) \in Add\_Des$}
			\State $Cand \gets Cand \cup \{(\bm{\mu},\bm{\Sigma},\textbf{V},\textbf{W},prob\cdot prob_{prev},1)\}$
			\EndFor
			\EndFor
			\For {$hor \in \{1,...,max\_horizon\}$}
			\For{$(\bm{\mu},\bm{\Sigma},prob_{prev}) \in Particles(t-hor)$}
			\State $\tilde{\textbf{Y}} \gets \left[\textbf{Y}_{t-hor+1}^T,...,\textbf{Y}_{t}^T\right]^T$
			\State $Inn\_Des \gets \text{BS\_inn}(\bm{\mu},\bm{\Sigma},\tilde{\textbf{Y}},\textbf{A},\textbf{C},\bm{\Sigma}_A,\bm{\Sigma}_I,M,hor)$
			\For {$(\textbf{V},\textbf{W},prob) \in Inn\_Des$}
			\State $Cand \gets Cand \cup \{(\bm{\mu},\bm{\Sigma},\textbf{V},\textbf{W},prob\cdot prob_{prev},hor)\}$
			\EndFor 
			\EndFor
			\EndFor
			\State $Desc \gets \text{Subsample}(N,Cand)$
			\Comment{Sampling proportional to $prob$}
			\State $Particles(t) \gets \{\}$
			\For{$(\bm{\mu},\bm{\Sigma},\textbf{V},\textbf{W},prob,hor) \in Desc $}
			\State $(\bm{\mu},\bm{\Sigma}) \gets  \text{KF\_Upd}(\textbf{Y}_{t+1-hor},\bm{\mu},\bm{\Sigma},\textbf{C},\textbf{A},\textbf{V}^{1/2}\bm{\Sigma}_A,\textbf{W}^{1/2}\bm{\Sigma}_I)$
			\If{$hor > 1$}
			\For {$i \in \{2,...,hor\}$}
			\State $(\bm{\mu},\bm{\Sigma}) \gets  \text{KF\_Upd}(\textbf{Y}_{t+i-hor},\bm{\mu},\bm{\Sigma},\textbf{C},\textbf{A},\bm{\Sigma}_A,\bm{\Sigma}_I)$
			\EndFor
			\EndIf
			\State $Particles(t) \gets Particles(t) \cup \{ (\bm{\mu},\bm{\Sigma},prob \cdot \frac{|Cand|}{|Desc|}) \}$
			\EndFor
			\EndFor
		\end{algorithmic}
	\end{footnotesize}
\end{algorithm}



\end{document}